\documentclass[english,11pt]{article}

\pdfoutput=1

\usepackage{amsmath}
\usepackage{amsfonts}
\usepackage{amssymb}
\usepackage{xfrac}
\usepackage{graphicx, rotating}
\usepackage{epstopdf}
\usepackage{epsfig}
\usepackage{latexsym}
\usepackage{multirow}
\usepackage[dvipsnames]{xcolor}
\usepackage{slashed}
\usepackage[top=2.8 cm, bottom=2.8 cm, left=2.5 cm, right=2.5 cm]{geometry}
\usepackage{amstext} 
\usepackage{array}  
\usepackage{hyperref}
\hypersetup{
	colorlinks = true,
	linkcolor = {Salmon},
	citecolor= {SeaGreen},
	urlcolor= {MidnightBlue}
}
\usepackage{braket}
\usepackage{dsfont}
\usepackage[compat=1.0.0]{tikz-feynman}
\usepackage{comment}
\usepackage{scalerel}
\usepackage{enumerate}
\usepackage{appendix}
\usepackage{tabularx}
\usepackage{booktabs}
\usepackage{multirow}
\usepackage{caption}
\usepackage{subcaption}
\usepackage{bm}
\usepackage[nottoc]{tocbibind}
\usepackage{cite}

\numberwithin{equation}{section}

\newcommand{\inv}{^{-1}}

\renewcommand{\Re}{\text{Re}}
\renewcommand{\Im}{\text{Im}}
\renewcommand{\phi}{\varphi}
\newcommand{\mul}{_{\mu}}
\newcommand{\nul}{_{\nu}}
\newcommand{\muh}{^{\mu}}

\newcommand{\sigmunu}{\sigma^{\mu\nu}}
\newcommand{\Lagr}{\mathcal{L}}
\newcommand{\eps}{\epsilon}
\newcommand{\gaC}{\gamma ^5}

\usepackage{lipsum}

\newcommand{ \mysmall}[1]{\scriptscriptstyle #1} 

\begin{document}
\unitlength = 1mm

\thispagestyle{empty} 
\begin{flushright}
\begin{tabular}{l}
\end{tabular}
\end{flushright}
\begin{center}
\vskip 3.4cm\par
{\par\centering \textbf{\LARGE High-energy frontier of the muon g-2\\at a muon collider}}

\vskip 1.2cm\par
{\scalebox{.85}{\par\centering \large  
\sc Paride Paradisi$\,^{a,b}$, Olcyr Sumensari$\,^{c}$, Alessandro Valenti$\,^{b,a}$}
{\par\centering \vskip 0.7 cm\par}
{\sl 
$^a$~Istituto Nazionale Fisica Nucleare, Sezione di Padova, I-35131 Padova, Italy}\\
{\par\centering \vskip 0.2 cm\par}
{\sl 
$^b$~Dipartamento di Fisica e Astronomia ``G.~Galilei", Università di Padova, Italy}\\
{\par\centering \vskip 0.2 cm\par}
{\sl 
$^c$~IJCLab, P\^ole Th\'eorie (B\^at.~210), CNRS/IN2P3 et Universit\'e Paris-Saclay, 91405 Orsay, France}\\

{\vskip 1.65cm\par}}
\end{center}

\vskip 0.85cm
\begin{abstract}

The long-standing muon g-2 anomaly can be explained by heavy new physics particles through chirally enhanced contributions.
It has been recently proposed that a muon collider running at center-of-mass energies of several TeV
could test these new physics scenarios in a model-independent way, through the study of 
high-energy processes such as $\mu^+ \mu^- \to h \gamma$. In this work, we validate these findings,
based on effective field theories, by considering selected renormalizable simplified models and by computing this one-loop process in full generality. 
Furthermore, we explore the interplay of direct and indirect high-energy searches to pin down the 
details of the underlying new physics model accommodating the muon g-2 anomaly.

\end{abstract}
\setcounter{page}{1}
\setcounter{footnote}{0}
\setcounter{equation}{0}
\noindent

\renewcommand{\thefootnote}{\arabic{footnote}}

\setcounter{footnote}{0}

\newpage

\section{Introduction}

The anomalous magnetic moment of the muon $a_\mu \!=\! (g_\mu \!-\! 2)/2$ represents one of the most interesting and long-standing hint for New Physics (NP). 
Recently, the E989 experiment at Fermilab~\cite{Abi:2021gix} has confirmed previous results by the E821 experiment at BNL~\cite{Bennett:2006fi}, yielding the 
experimental average $a_\mu^{\mysmall \rm EXP} \!=\! 116592061(41) \!\times\! 10^{-11}$. Comparing this value with the Standard Model (SM) prediction 
$a_\mu^{\mysmall \rm SM} \!=\! 116591810(43) \times 10^{-11}$, reported by the Muon $g$-2 Theory Initiative~\cite{Aoyama:2020ynm}, leads to an interesting $4.2\,\sigma$ discrepancy~\cite{Abi:2021gix}~\footnote{Recently, a lattice QCD collaboration computed the leading hadronic contribution to the muon $g$-2 with a comparable precision to the dispersive determinations, finding a larger value which weakens the discrepancy to $1.6\sigma$~\cite{Borsanyi:2020mff}. However, this increase to the hadronic contribution could imply tensions with the electroweak fit, or with low-energy $e^+e^- \!\to\! {\rm hadron}$ data~\cite{Passera:2008jk}. For this reason, the findings of Ref.~\cite{Borsanyi:2020mff} should be verified by independent lattice QCD studies which are underway
or by direct experimental measurements, as proposed by the MUonE experiment~\cite{CarloniCalame:2015obs}.} 
\begin{equation}
\Delta a_\mu = a_\mu^{\mysmall \rm EXP}-a_\mu^{\mysmall \rm SM} = 251 \, (59) \times 10^{-11}\,.
\label{eq:gmu}
\end{equation}

\noindent Since the observed deviation is comparable in size to the SM electroweak contribution, it would be natural to invoke new weakly-coupled particles at the electroweak scale
to solve this puzzle. However, this possibility is strongly disfavoured by LEP and LHC data which push the NP scale $\Lambda$ to lie above $\Lambda \gtrsim 1~$TeV.~\footnote{Other viable solutions are provided by very light and feebly coupled NP particles such as axionlike particles~\cite{Marciano:2016yhf}.} 

Heavy NP contributions to $\Delta a_\mu$ are captured by the dimension-6 operator $\left(\bar\mu_L \sigma_{\mu\nu} \mu_R\right) H F^{\mu\nu}$~\cite{Buchmuller:1985jz}, 
where $H$ is the SM Higgs doublet and $F^{\mu\nu}$ denotes the electromagnetic field strength tensor.
After electroweak symmetry breaking, $\Delta a_\mu$ receives the contribution $\Delta a_\mu \sim ({g^3} _{\mysmall\rm NP}/ 16\pi^2) \times (m_\mu v/\Lambda^2)$, 
where $v=246$~GeV is the electroweak vacuum-expectation-value (vev) and $g_{\rm\mysmall NP}$ denotes a generic NP coupling. Therefore, the NP chiral enhancement 
$v/m_\mu \sim 10^3$ brings the sensitivity of $\Delta a \mul$ to NP scales of order $\Lambda \sim 10\,$TeV even for weak couplings $g_{\rm\mysmall NP} \sim 1$~\cite{Giudice:2012ms,Capdevilla:2020qel}.
The same dipole operator generating $\Delta a_\mu$ induces also a NP contribution to the process $\mu^+\mu^- \to h \gamma$ that grows quadratically with the center-of-mass energy  
$\sqrt{s}$ of the collisions, as recently demonstrated in the context of Effective Field Theories (EFT)~\cite{Buttazzo:2020ibd}. 
Therefore, measuring the cross section of $\mu^+\mu^- \to h \gamma$ would be equivalent to measuring $\Delta a_\mu$. This goal can be achieved at 
a multi-TeV muon collider~\cite{Delahaye:2019omf}.

In this work, we revisit the connection between $\Delta a_\mu$ and $\mu^+ \mu^- \to h \gamma$ within simplified models which induce chirally enhanced contributions to $\Delta a_\mu$. 
In particular, we focus on models with new scalars and vectorlike fermions in various $SU(2)_L\times U(1)_Y$ representations, with an underlying $Z_2$ symmetry to prevent dangerous mixing of the new states with SM fields~\cite{Crivellin:2021rbq,Arcadi:2021cwg}. 
As already discussed in Ref.~\cite{Crivellin:2021rbq}, where the matching of these models onto the relevant set of dimension-$6$ SMEFT operators~\cite{Buchmuller:1985jz} 
has been performed, these scenarios display correlations between $\Delta a_\mu$ and other processes such as $h\to\mu^+ \mu^-$ and $Z\to\mu^+ \mu^-$.
Moreover, such models generally contain a stable particle and therefore they can also explain the observed dark matter relic abundance~\cite{Kowalska:2017iqv,Calibbi:2018rzv}.

The first goal of our analysis is to validate the findings of previous EFT studies~\cite{Buttazzo:2020ibd} by performing a full one-loop calculation of the $\mu^+ \mu^- \to h \gamma$ cross section
within the simplified models of Ref.~\cite{Crivellin:2021rbq}. As our results hold for any center-of-mass energy value $\sqrt{s}$, they will complement the findings of Ref.~\cite{Buttazzo:2020ibd}, 
which only apply in the EFT regime $\sqrt{s} \ll \Lambda$, and they will allow us to precisely assess the validity limit of the EFT description for this particular process.
Another goal of our work is to study the direct searches signatures of these simplified models (see also Ref.~\cite{Capdevilla:2020qel}). On general grounds, the discovery of new particles by their direct production can be hardly associated 
in a nonambiguous way to $\Delta a_\mu$. However, this statement strictly applies only to $2\to 2$ processes as they are not sensitive to the same combination
of parameters entering $\Delta a_\mu$. Instead, we point out that $2\to 3$ processes with a Higgs boson in the final state exhibit a stronger correlation with $\Delta a_\mu$ and  with $\mu^+\mu^- \to h \gamma$. The correlated study of these observables at a muon collider may allow to disentangle among the underlying NP model accommodating the 
$\Delta a_\mu$ anomaly, therefore representing a very interesting example of the interplay of the high-energy and high-intensity frontiers of particle physics. 
 
The paper is organized as follows. In section II, we introduce the simplified models and their predictions for the muon $g$-2.
In section~III, we focus on indirect high-energy probes of the muon $g$-2 at a muon collider, by computing the one-loop induced process $\mu^+\mu^- \to h\gamma$
in the context of simplified models and by establishing the limit of validity of the EFT results.
In section~IV, we analyze direct high-energy probes of the muon $g$-2 at a muon collider which include both $2\to 2$ and $2\to 3$ scattering processes.
Our final remarks and conclusions are made in Sec.~V.

\section{Simplified models for the muon \boldmath{$g$-2}}
\label{sec:models}

We consider the two classes of simplified models that can provide a chiral enhancement to $\Delta a_\mu$, which consist in extending the SM with two scalars $\Phi_{L,E}$ and one vectorlike fermion $\Psi$ (model I), or two vectorlike fermions  $\Psi_{L,E}$ and one scalar $\Phi$ (model II). These models are generically described by the following Lagrangians~\cite{Crivellin:2021rbq,Calibbi:2018rzv},~\footnote{The quartic couplings between the SM Higgs and the new scalars are not explicitly written since they are irrelevant for our phenomenological study.}
\begin{align}
{\Lagr_\text{I}} &= \lambda _L^{\text{I}}\,\bar \ell\Psi {\Phi _L} + \lambda _E^{\text{I}}\,\bar e\Psi {\Phi _E} + A\,\Phi _L^\dag \Phi _E^{} \, H +\mathrm{h.c.}\,, 
\\[0.35em]
{\Lagr_\text{II}} &= \lambda _L^\text{II}\,\bar \ell{\Psi _L}\Phi  + \lambda _E^\text{II}\,\bar e{\Psi _E}\Phi  + \kappa\, {{\bar \Psi }_L}{\Psi _E}H +\mathrm{h.c.}\,,
\label{eq:simplified_models}
\end{align}
where $\ell$ and $e$ are the SM lepton doublet and singlet, respectively, and $H$ denotes the SM Higgs doublet. Note, in particular, that we have imposed an underlying $Z_2$ symmetry to prevent dangerous mixing of the new states with SM fields~\cite{Crivellin:2021rbq}.
By restricting the $SU(2)_L$ representations of $\Psi_{(L,E)}$ and $\Phi_{(L,E)}$ up to triplets, there are four possibilities in each of these models. The allowed $SU(2)_L\times U(1)_Y$ representations are listed in Table~\ref{tab:s1charges}, where  $X$ denotes the hypercharge of the field $\Psi$ ($\Phi$) for the models of type I (type II). The respective Lagrangians are spelled out in Appendix~\ref{app:running} where the $SU(2)_L$ contractions are explicitly written.~\footnote{The new particles could also be charged under $SU(3)_c$, which would amount to multiplicative representation-dependent factors in the expressions derived in this paper. In particular the fermions $\Psi_{L,E}$ in model II could be the top-quark, recovering the minimal leptoquark solution to  $\Delta a_\mu$ where $\Phi$ could be either the state $(3,2,7/6)$ or $(\bar{3},1,1/3)$~\cite{Cheung:2001ip}. Instead, scenarios with two scalar leptoquarks are fully described by model I upon matching Ref.~\cite{Dorsner:2019itg}.}

\begin{table}[!htb]
	\centering
	\begin{tabular}{cc|ccc}
		& $R$ & ${\Psi ,\Phi }$&${{\Phi _L},{\Psi _L}}$&${{\Phi _E},{\Psi _E}}$\\
		\midrule 
		\multirow{4}{*}{$SU(2)_L$} & $121$ & $1$ & $2$ & $1$ \\
		& $212$ & $2$ & $1$ & $2$ \\
		& $323$ & $3$ & $2$ & $3$ \\
		& $232$ & $2$ & $3$ & $2$ \\
		$Y$ & & $X$ & $-\frac{1}{2}-X$ & $-1-X$\\
		\bottomrule
	\end{tabular}
	\caption{Charge assignments and representations under $SU(2)_L\times U(1)_Y$ for the NP states.}
	\label{tab:s1charges}
\end{table}

\begin{figure}[t!]
	\centering
	 \includegraphics[width = 0.85\textwidth]{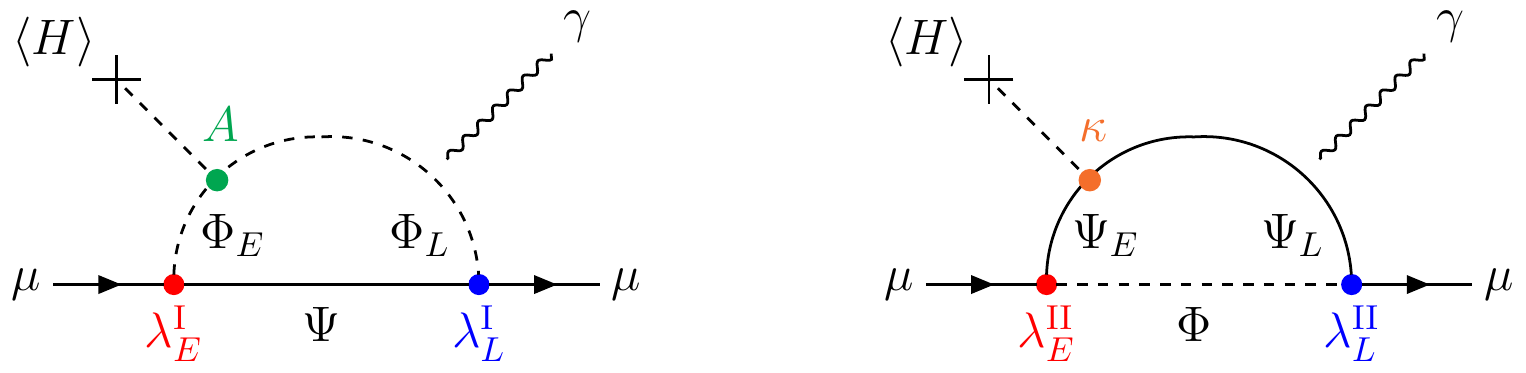}
	\caption{\label{fig:Feynman_I} Feynman diagrams contributing to the muon $g$-2 at one-loop level in the simplified models I (left panel) and II (right panel).}
\end{figure}

The simplified models listed in Table~\ref{tab:s1charges} contribute to $\Delta a_\mu$ via the loop diagrams depicted in Fig.~\ref{fig:Feynman_I}. These new interactions are assumed to arise at the energy scale $\Lambda$, which lies well above the electroweak scale, in such a way that their contributions to $\Delta a_\mu$ can be fully interpreted in terms of the SMEFT Lagrangian~\cite{Buchmuller:1985jz}, 
\begin{equation}
\label{eq:eft-gminus2}
\mathcal{L}_\mathrm{SMEFT} \supset {C_{eB}} \,\big{(}\bar{\ell} \sigma^{\mu\nu} e\big{)}H B_{\mu\nu} + {C_{eW}} \,\big{(}\bar{\ell} \sigma^{\mu\nu} e\big{)} \tau^I H\, W^I_{\mu\nu} +\mathrm{h.c.}
\end{equation}

\noindent where we have only written the $d=6$ operators that are relevant in our setup, and where flavor indices are omitted. The leading contributions to $\Delta a_\ell$ can then be written as
\begin{equation}
\Delta a_\ell \simeq \dfrac{4 m_\ell v}{\sqrt{2} e } \mathrm{Re}\left(C_{e\gamma}\right)\,,
\end{equation}

\noindent where the effective coefficient $\smash{C_{e\gamma}^\ell= \cos \theta_W\, C_{eB}^\ell- \sin \theta_W\, C_{eW}^\ell}$ can be expressed, for the simplified models I and II, in a very compact form %
\begin{align}
\label{eq:Deltaamu_I}
\big{[}C_{e\gamma}\big{]}_\mathrm{I} &= - \dfrac{e\Re[\lambda_L^\text{I}(\lambda_E^\text{I})^* A]}{384 \pi^2  M^3} \Big\{2X+1,\,-2X\,,6X-1\,,2(3X+2)\Big\}\,, \\[0.35em]
\label{eq:Deltaamu_II}
\big{[}C_{e\gamma}\big{]}_\mathrm{II} &= \dfrac{e\Re[\lambda_L^\text{II}(\lambda_E^\text{II})^* \kappa]}{384 \pi^2 M^2}  \Big\{2(X+1),\,-(2X+1),\,2(3X+1),\,6X+7\Big\}\,,
\end{align}
where a degenerate mass $M$ is assumed for the NP states. The expressions between brackets give the hypercharge factors for the $SU(2)_L$ representations $R=\lbrace 121, 212, 323, 232\rbrace$ of Table~\ref{tab:s1charges}.
Our results shown in Eqs.~(\ref{eq:Deltaamu_I}) and (\ref{eq:Deltaamu_II}) are in full agreement with those from Ref.~\cite{Crivellin:2021rbq}.

Since we are interested in scenarios with new particles in the multi-TeV range, an explanation of the muon $g$-2 anomaly would necessarily require $\mathcal{O}(1)$ Yukawa couplings. 
In particular, one typically finds that a contribution of order $\Delta a_\mu \sim 10^{-9}$ can be obtained for $\lambda_{L}, \lambda_{E}, \kappa \sim 2$, $A/M \sim 1$ and $M \sim 10$ TeV. Two main concerns may arise for such large couplings. First, the very same chiral enhancement in $\Delta a\mul$ is also at work in the quantum corrections to the muon mass. Therefore, a new naturalness problem involving the muon mass is typically present in these models, see for instance Ref.~\cite{Capdevilla:2020qel} where a careful analysis of the parameter space of these models has been performed. Secondly, for such large couplings the quantum stability of the simplified models should be carefully checked, as we do in the following.

\begin{figure}[!h]
	\centering
	\includegraphics[width = 0.32\textwidth]{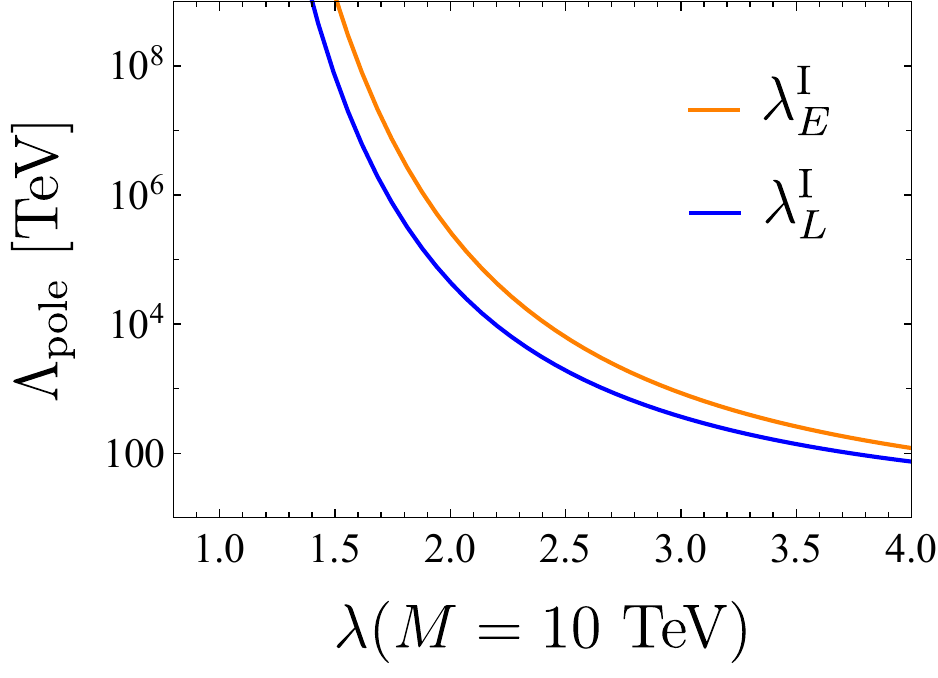}
	\includegraphics[width = 0.32\textwidth]{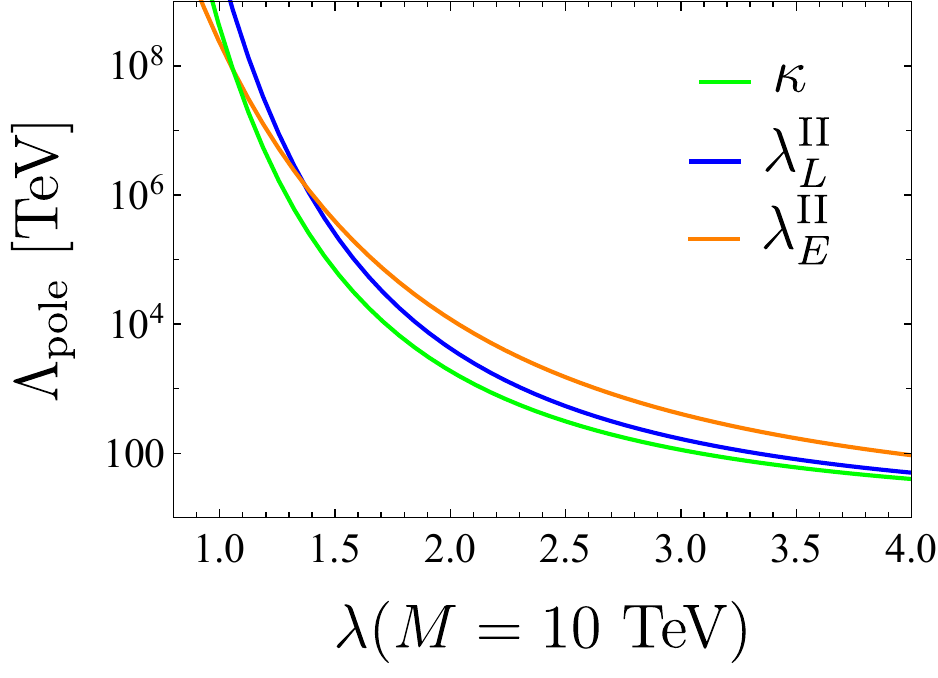}	
	\includegraphics[width = 0.32\textwidth]{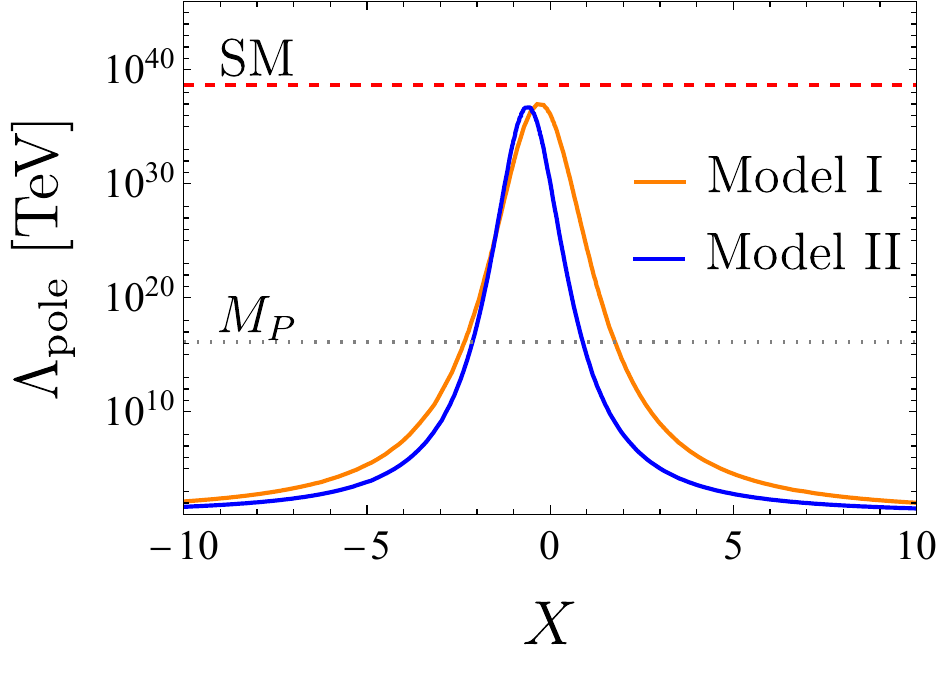}
	\caption{Landau poles of the new coupling constants of the simplified models I (left) and II (center) for the representation $R = 121$ (see Appendix \ref{app:running}) setting $M=10$ TeV and $X=1/2$. 
	The plot on the right shows the location of the Landau pole of the SM gauge coupling $g'$ as a function of X.}
	\label{fig:Landau_pole}	
\end{figure}

The one-loop running of the model parameters for the various choices of the $SU(2)_L\times U(1)_Y$ representations is reported in detail in Appendix \ref{app:running}.
As a result of this analysis, we show in Fig.~\ref{fig:Landau_pole} (left and center panels) the Landau poles of the new coupling constants $\lambda_L$, $\lambda_E$ and $\kappa$ for the representation 
$R = 121$ setting $M=10$ TeV and $X=1/2$.~\footnote{A similar analysis has been performed in Ref.~\cite{Capdevilla:2020qel}, taking into account the NP contributions to the running of the SM parameters. In addition to these effects, we also account for the running of the NP couplings that can develop Landau Poles even before the SM ones.} By requiring that these couplings do not develop a Landau pole below $10^3~\mathrm{TeV}$, we conclude that they should be smaller than $\approx 3$. In the right panel, we also plot the location of the Landau pole of the SM gauge coupling $g'$ as a function of the hypercharge $X$. Clearly, the absolute value of $X$ cannot be arbitrarily large, otherwise $g^\prime$ would develop a pole well below the Planck scale. By combining these indirect bounds, we infer that the simplified scenarios can only be self-consistent, while explaining explain the $\Delta a_\mu$ discrepancy, if the mass $M$ is below $\lesssim 15~\mathrm{TeV}$. These conclusions have been obtained for the representation $R=121$ and $X=1/2$, but they can be easily generalized to the other scenarios. Note, in particular, that in the presence of weak triplets, one should also worry about the Landau poles of the $SU(2)_L$ gauge coupling $g$ since its $\beta$-function becomes positive for some of the models from Table~\ref{tab:s1charges}.


%

\section{Indirect high-energy probes of the muon \boldmath{$g-2$}}
\label{sec:xsect}

\begin{figure}[!htb]
	\centering
	\includegraphics[width = 0.65\textwidth]{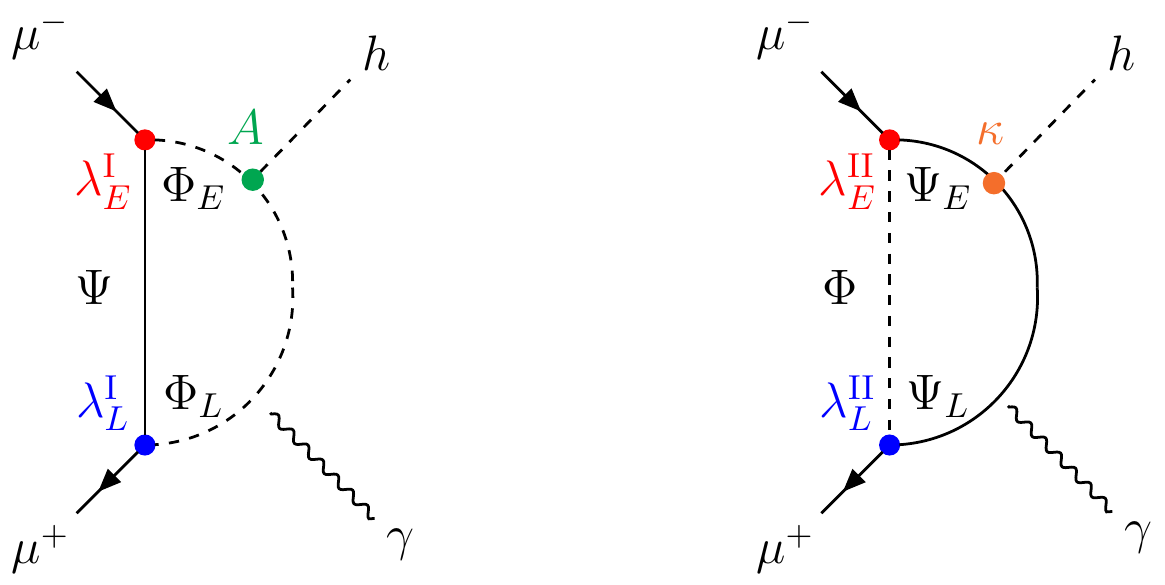}
	\caption{\label{fig:Feynman_II} Feynman diagrams contributing to the high-energy scattering $\mu^+ \mu^- \rightarrow h \gamma$
	for the simplified models I (left panel) and II (right panel).}
\end{figure}

As recently discussed in Ref.~\cite{Buttazzo:2020ibd}, an interesting signature which inevitably accompanies NP contributions to $\Delta a_{\mu}$ is the modification to the process $\mu^+ \mu^- \rightarrow h \gamma$ at high-energies, which could be a target of the proposed muon collider~\cite{Delahaye:2019omf}. This complementarity becomes clear when comparing the Feynman diagrams depicted in Figs.~\ref{fig:Feynman_I} and \ref{fig:Feynman_II}.
In this section, we will compute the cross section of $\mu^+ \mu^- \rightarrow h\gamma$ in the simplified models outlined above. Since our results will be valid for any center-of-mass energy value $\sqrt{s}$, 
they will complement the results of Ref.~\cite{Buttazzo:2020ibd} which only apply in the EFT regime $\sqrt{s}\ll \Lambda$. As a by-product, the comparison between the computations made within concrete scenarios and the corresponding EFTs will allow us to precisely assess the limits of the EFT description for this particular process.

We start by writing the most general Lorentz-invariant amplitude contributing to $\mu^+ \mu^- \rightarrow h\gamma$,
\begin{align}
\begin{split}
	\mathcal{A}_\mu = \bar v (p_2) &\left[ A \gamma_\mu + B (p_1 + p_2)_\mu + C k_\mu + D\sigma_{\mu\nu} k^\nu + 
E \sigma_{\mu\nu} (p_1+p_2)^\nu \right.  \\
&+ \left. A^\prime \gamma_\mu \gamma_5 + B^\prime \gaC (p_1 + p_2)_\mu + C^\prime \gaC k_\mu + D^\prime\gaC \sigma_{\mu\nu}  k^\nu + 
E^\prime \gaC \sigma_{\mu\nu} (p_1+p_2)^\nu \right] u(p_1)\,,
	\label{eq:s3FormFactors}
\end{split}
\end{align}
where $p_1$, $p_2$ and $k$ denotes the momentum of $\mu^-$, $\mu^+$ and $\gamma$, respectively. In this expression we exploited momentum conservation, $q\muh = p_1 \muh +p_2 \muh -k\muh$, and neglected the masses of the external states. The form factors $\lbrace A^{(\prime)}, B^{(\prime)},\dots \rbrace$ are functions of the 
Mandelstam variables $s$, $t$ and $u$ defined as follows,
\begin{align}
\label{eq:stu}
s=(p_1+p_2)^2\,,\qquad\quad t=(p_1-k)^2\,, \qquad\quad u=(p_2-k)^2\,,
\end{align}
%
which satisfy $s+t+u\approx 0$. The expression of Eq.~(\ref{eq:s3FormFactors}) has been obtained taking into account that:
\begin{itemize}
	\item The Gordon's identity and the Dirac equation allow to trade $\sigmunu(p_1 \mp p_2) \nul$ for $(p_1 \pm p_2) \muh$  in the massless muon limit. 
	The same holds for the corresponding terms with an extra $\gaC$;
	\item A possible term $\bar v\,\eps_{\mu \nu \rho \sigma} p_{1}^{\nu} p_{2}^{\rho}k^{\sigma} u$ containing the Levi-Civita tensor $\eps_{\mu\nu\rho\sigma}$ is not independent. 
	 In fact, we can write $\epsilon_{\mu\nu\rho\sigma} = -i \gaC \gamma_{[\mu} \gamma_{\nu}\gamma_{\rho}\gamma_{\sigma]}$, reducing this term to
	 $\bar v \,\gamma \muh \slashed{p_1}\slashed{p_2}\slashed{k}\, u+ \mbox{permutations}$. Then, by using $\slashed{a}\slashed{b}= a\cdot b-i \sigmunu a\mul b \nul$, we are able to recast this term as a linear combination of the ones already listed in Eq.~(\ref{eq:s3FormFactors}). 
	 The same is true for the term $\bar v\gamma_5\eps_{\mu \nu \rho \sigma} p_{1}^{\nu} p_{2}^{\rho}k^{\sigma}u$\,.
\end{itemize}	 
Furthermore, Eq.~(\ref{eq:s3FormFactors}) can be simplified by imposing that the amplitude $\mathcal{A}_\mu$ is gauge invariant, that is imposing the QED Ward identity $k^\mu\mathcal{A}_\mu=0$. As a result, we find that $A=A^\prime=0$, $B^\prime = - i E^\prime$ and $s B= (u-t)iE$. 
Finally, dropping the irrelevant $k_\mu$ term which gives a vanishing contribution for on-shell photons, we can write the amplitude in a very compact form
\begin{align}
\begin{split}
	\mathcal{A}\muh= \bar v (p_2) 
	&\bigg[ \mathcal{D}\, i\sigmunu k\nul  + \mathcal{F} \,\bigg(\frac{t-u}{s}(p_1+p_2)\muh -\frac{t+u}{s}(p_1-p_2)\muh\bigg)
	\\
	& +\mathcal{D}^\prime\,i\sigmunu \gamma_5 k\nul  + \mathcal{F}^\prime\, \bigg( \dfrac{t-u}{s}(p_1+p_2)\muh- \dfrac{t+u}{s}(p_1-p_2)\muh  \bigg) \gamma_5 
	\bigg] u(p_1)\,,
	\label{eq:FF}
\end{split}
\end{align}
where $\mathcal{D}^{(\prime)}$ and $\mathcal{F}^{(\prime)}$ are the only independent form-factors, which are defined as linear combinations of the ones defined above. The $\mu^+\mu^-\to h\gamma$ differential cross section can then be written as
\begin{align}
	\frac{	d\sigma_{h\gamma}}{dt} = \frac{tu}{16\pi s^2} 
	\left(
	\left|\mathcal{D}+\mathcal{F}\right|^2 +  \left|\mathcal{F}\right|^2 +
	\left|\mathcal{D'}+\mathcal{F'}\right|^2 + \left|\mathcal{F'}\right|^2
	\right)\,.
\end{align}
We are now ready to calculate the analytical expressions of the form factors $\mathcal{D}^{(\prime)}$ and $\mathcal{F}^{(\prime)}$ in the simplified models I and II.

Our approach is to evaluate the amplitudes associated with the Feynman diagrams of Fig.~\ref{fig:Feynman_II} and then to project them into the form factors of Eq.~(\ref{eq:FF}). The results of our full computation will be presented in Sec.~\ref{ssec:full}. The discussion fo the EFT limit for the form-factors will be made in Sec.~\ref{ssec:eft} and our numerical results will be presented in Sec.~\ref{ssec:numerical}.

\subsection{\boldmath{$\mu^+\mu^-\to h \gamma$} in simplified models}
\label{ssec:full}

In this section, we explicitly evaluate the form-factors $\mathcal{D^{(\prime)}}$ and $\mathcal{F^{(\prime)}}$ at one-loop for the simplified models defined in Eq.~\eqref{eq:simplified_models}. Our convention for the kinematical variables is given in Eq.~\eqref{eq:stu} and we use {\tt Package-X}~\cite{Patel:2016fam} to reduce the one-loop integrals in terms of the Passarino-Veltman functions. The masses of the external states are neglected and, for simplicity, we assume a degenerate mass $M$ for the new scalar and vectorlike fermion running in the loops. {Although we consider the case of degenerate scalar and fermion masses, we provide a {\small \sc Mathematica} notebook in the ancillary files of this paper with expressions that also hold in the case of nondegenerate masses. In particular, we note that the cross section can be increased/decreased for $\sqrt s$ values above the lightest mass. In any case, for energies below the lightest mass threshold, we stress that the EFT predictions should be reproduced irrespectively of the degenerate or nondegenerate case.}

In order to obtain general results for all the models appearing in Table~\ref{tab:s1charges}, we define the coefficients $\xi$ and $\tilde{\xi}$ in Table \ref{tab:s3repFactors} which depend on the $SU(2)_L$ representation $R$ that is considered.~\footnote{These coefficients can be compared to Ref.~\cite{Crivellin:2021rbq} by identifying $\xi \leftrightarrow \xi_{eB}$, $\tilde \xi \leftrightarrow \tilde \xi_{eW}$ and noting that their extra two coefficients are not independent, i.e.~$\xi_{eW}^E = -\tilde \xi_{eW}, \xi_{eW}^L= -\tilde \xi_{eW}+\xi_{eB}/2$. The last relations follow from gauge invariance and can be obtained, for instance, by explicitly checking the Ward identity in the $ \mu ^+\mu ^- \rightarrow h\gamma$ amplitude.} 
To express the amplitude in terms of physical parameters, we define in a first step the physical muon Yukawa coupling as the sum of the tree-level coupling and the 1-loop corrections. Then, we write the $\mu^+\mu^-\rightarrow h\gamma$ amplitude as the sum of the contributions from the diagrams of Fig.~\ref{fig:Feynman_II} and the tree-level diagrams where the bare muon Yukawa-coupling is replaced with the physical Yukawa-coupling and the loop corrections computed in the first step. This renormalization procedure automatically removes all the possible divergences to the $\mu^+ \mu^- \rightarrow h \gamma$ amplitude.

\begin{table}[!htb]
	\centering
	\begin{tabular}{crrrr}
		\toprule
		$R$& $121$ & $212$ & $323$ & $232$\\
		\midrule
		$\xi$ & $1$ & $-1$ & $3$ & $3$\\
		$\tilde\xi$ & $0$ & $-\frac{1}{2}$ & $2$ & $-\frac{1}{2}$\\
		\bottomrule
	\end{tabular}
	\caption{Representation-dependent factors $\xi$ entering the $\mu^+\mu^- \to h \gamma$ calculation for each of the models defined in Table~\ref{tab:s1charges}.}
	\label{tab:s3repFactors} 
\end{table}
%

\paragraph{\underline{Model I}:} For the simplified model I, we find the following results,
\begin{align}
\mathcal{D} &= \frac{ ieM\Re[\lambda_L^\text{I}(\lambda_E^\text{I})^* A] }{32\sqrt{2}\pi^2  t u}\left[ 2(\xi X- \tilde \xi) t u D_0(0,0,0,0,t,u;\mathbf{M})-2\xi u C_0(0,0,t;\mathbf{M}) \right. \nonumber \\
	& \quad  \left.-  \xi t C_0(0,0,u;\mathbf{M})+\xi \frac{s}{M^2}\right]\,, \\[0.55em]
\mathcal{F}&= -\frac{ieM \Re[\lambda_L^\text{I}(\lambda_E^\text{I})^* A]}{ 32  \sqrt{2} \pi^2 t u} \left[ \left(\xi(1+X)-\tilde \xi\right) su D_0(0,0,0,0,s,u;\mathbf{M}) \right.  +tu (\xi X-\tilde \xi) D_0(0,0,0,0,t,u;\mathbf{M}) \nonumber \\
	 &\quad+  s t \left(\xi(1+X)-\tilde \xi\right)   D_0(0,0,0,0,s,t;\mathbf{M})+2 \left(\xi s+(\tilde\xi-\xi X)u \right) C_0(0,0,u;\mathbf{M})\nonumber \\
	 &\quad+2 \left(\xi s+(\tilde\xi-\xi X)t \right) C_0(0,0,t;\mathbf{M})
	 \left.-  2\left(\xi(1+X)-\tilde \xi\right)sC_0(0,0,s;\mathbf{M})-\xi\frac{s}{M^2} \right]\,,
\end{align}
where $C_0$ and $D_0$ are scalar Passarino-Veltman functions, and we adopt the notation $\mathbf{M} \equiv (M,M,M)$ and $\mathbf{M} \equiv (M,M,M,M)$ in the arguments of $C_0$ and $D_0$, respectively. The convention on the arguments of the scalar functions follows Ref.~\cite{Patel:2016fam}. The form factors $\mathcal{F}^ \prime$ and $\mathcal{D}^\prime$ are simply obtained from the above expressions upon the substitution $\Re[\lambda_L^\text{I}(\lambda_E^\text{I})^* A] \rightarrow i\Im[\lambda_L^\text{I}(\lambda_E^\text{I})^* A]$.

In order to compare the above results with those obtained with the EFT approach (see Sec.~\ref{ssec:eft}), we perform a power expansion in $\sqrt{s}/M \ll 1$, 
\begin{align}	 
\label{eq:taylor-Ia}
	\mathcal{D} &\simeq \frac{i e \Re[\lambda_L^\text{I}(\lambda_E^\text{I})^* A]}{192 	\sqrt{2}\pi^2 M^3} \left\{\left[\xi(1+2X)-2\tilde\xi\right]- \frac{1}{15} \left[\xi(1+3X)-3\tilde\xi\right]\frac{s}{M^2} \right. \\ 
	& \qquad\qquad\left. +\frac{1}{280} \left[\left(\xi(3+8X)-8\tilde\xi \right)-6\frac{tu}{s^2}\left(\xi(1+2X)-2\tilde\xi\right)\right]\frac{s^2}{M^4}+\mathcal{O}\left( \frac{s^3}{M^6}\right)\right\}\,, \nonumber\\[0.55em]
\label{eq:taylor-Ib}
	\mathcal{F} & \simeq \frac{i e  \Re[\lambda_L^\text{I}(\lambda_E^\text{I})^* A]}{192 	\sqrt{2} \pi^2 M^3} \left\{ \frac{1}{30}\left[\xi(2+3X)-3\tilde\xi\right]\frac{s}{M^2} + \frac{\xi}{280}\frac{s^2}{M^4}  +\mathcal{O}\left( \frac{s^3}{M^6}\right)\right\}\,.
\end{align}
\noindent A few comments on the above expressions are in order: $i)$ at $d=6$ level, only the first term of the form factor $\mathcal{D^{(')}}$ survives and it precisely reproduces the EFT result, see Eq.~\eqref{eq:ff-eft-D} in the following section; $ii)$ higher order terms are highly suppressed by small numerical coefficients making the EFT result quite accurate even for $\sqrt{s} \sim M$\,. This finding is rather unexpected, since in most cases the breakdown of the EFT description quickly arises as the energy $\sqrt{s}$ approaches the EFT cutoff.

 \paragraph{\underline{Model II}:}  In the case of the simplified model II, we follow the same procedure outlined above. The resulting analytical expressions for the form factors read,
\begin{align}
	\mathcal{D}=& \frac{i  e M^2 \Re[\lambda_L^\text{II}(\lambda_E^\text{II})^* \kappa]}{32 \sqrt{2}\pi^2 s t u} \left[-2 u \left(2  \xi s+(\xi X-\tilde\xi)\frac{t^2}{M^2} \right) C_0(0,0,t;\mathbf{M}) \right. \nonumber \\
	& -2 t \left(2  \xi s+(\xi X-\tilde\xi)\frac{u^2}{M^2}\right) C_0(0,0,u;\mathbf{M}) -2 \xi \frac{s t}{M^2} \Lambda
	(u,\mathbf{M})-2 \xi \frac{s u}{M^2} \Lambda (t;\mathbf{M}) \nonumber \\
	&\left. +(\xi X-\tilde\xi) tu  \left(4s+\frac{t u}{M^2}\right) D_0(0,0,0,0,t,u;\mathbf{M} )+6 \xi \frac{s^2}{M^2}\right]\,,\\[0.5em]
	\mathcal{F}=& -\frac{ie M^2\Re[\lambda_L^\text{II}(\lambda_E^\text{II})^* \kappa]}{32
		\sqrt{2} \pi^2 s t u} \left[-4 s^2 \left(\xi(1+X)-\tilde \xi\right) C_0(0,0,s;\mathbf{M}) \right. \nonumber \\
		&+2 \left(2  s (\xi s-(\xi X-\tilde\xi)t )-\frac{ut^2}{M^2} X \right) C_0(0,0,t;\mathbf{M})  \nonumber \\
	&+2 \left(2  s (\xi s-(\xi X-\tilde\xi)u )-\frac{u^2 t}{M^2} X \right) C_0(0,0,u;\mathbf{M}) -2 \left(\xi(1+X)-\tilde \xi\right)s t^2 
	D_0(0,0,0,0,s,t;\mathbf{M}) \nonumber \\
	&-2 \left(\xi(1+X)-\tilde \xi\right) s u^2   D_0(0,0,0,0,s,u;\mathbf{M})+ \left(\xi X-\tilde \xi\right)tu  \left( \frac{tu}{M^2}+4s\right)  D_0(0,0,0,0,t,u;\mathbf{M})\nonumber \\
	&-2\xi\frac{ s t}{M^2} \Lambda (u;\mathbf{M}) \left. -2 \xi \frac{su}{M^2} \Lambda (t;\mathbf{M})+6\xi\frac{s^2}{M^2}\right]\,,
\end{align}

\noindent where $\Lambda (x,M_1,M_2)$ is the part of the Passarino-Veltman $B_0$ function containing the $x$ plane branch cut~\cite{Patel:2016fam}. Again, the form factors for the $\gaC$ terms are identical upon the substitution $\Re[\lambda_L^\text{II}(\lambda_E^\text{II})^* \kappa] \rightarrow i\Im[\lambda_L^\text{II}(\lambda_E^\text{II})^* \kappa]$.

The low-energy expansion $\sqrt{s}/M \ll 1$ of the form-factors given above reads

\begin{align}	
\label{eq:taylor-IIa}
	\mathcal{D} \simeq& -\frac{ie \Re[\lambda_L^\text{II}(\lambda_E^\text{II})^* \kappa]}{96
		\sqrt{2} \pi^2 M^2}
	\left\{ \left[\xi(1+X)-\tilde \xi\right] - \frac{1}{60}\left[\xi(2+3X)-3\tilde \xi\right] \frac{s}{M^2} \right. \\
	& \left. + \frac{1}{840}\left[ (\xi(3+4X)-4\tilde \xi)-6(\xi(1+X)-\tilde \xi)\frac{tu}{s^2} \right]\frac{s^2}{M^4}  + \mathcal{O}\left( \frac{s^3}{M^6}\right)\right\},\nonumber \\
\label{eq:taylor-IIb}
	\mathcal{F} \simeq&  -\frac{ie  \Re[\lambda_L^\text{II}(\lambda_E^\text{II})^* \kappa]}{96\sqrt{2}\pi^2 M^2} 
	\left\{ \frac{1}{60}\left[\xi (8+9X)-9\tilde \xi\right]\frac{s}{M^2} +\frac{1}{840}\left[\xi (15+14X)-14\tilde \xi\right] \frac{s^2}{ M^4} +\mathcal{O}\left( \frac{s^3}{M^6}\right) \right\}.
\end{align}
As before, the EFT amplitude is correctly reproduced and the subleading power corrections are suppressed by large numerical factors.

\subsection{\boldmath{$\mu^+\mu^-\to h \gamma$} in EFT}
\label{ssec:eft}

Before presenting our numerical results, we remind the reader of the EFT description of the $\mu^+\mu^-\to h \gamma $ process. To this purpose, we assume that the center-of-mass energy $\sqrt{s}$ is much larger than the masses involved in this process, but still sufficiently smaller than the EFT cutoff. In this case, the $\mu^+\mu^- \to h \gamma $ scattering is dominated by a single $d=6$ operator, $\mathcal{O}_{e\gamma}=\bar{\ell} \sigma^{\mu\nu} e H F_{\mu\nu}$\,, which is a linear combination of the operators defined in Eq.~\eqref{eq:eft-gminus2}. 

\begin{figure}[!htb]
	\centering
	\includegraphics[width = 0.47\textwidth]{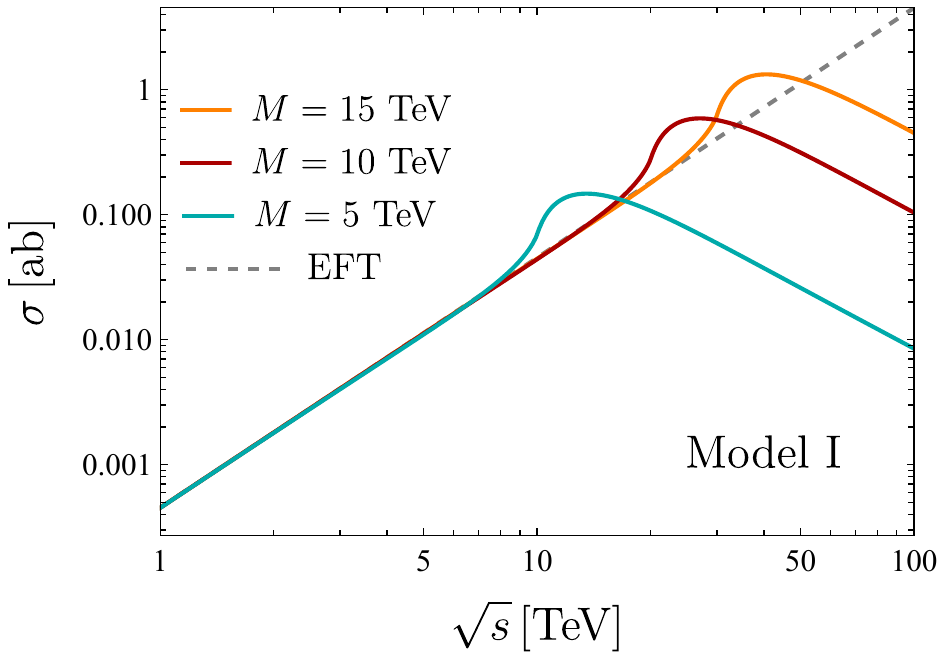}\qquad
	\includegraphics[width = 0.47\textwidth]{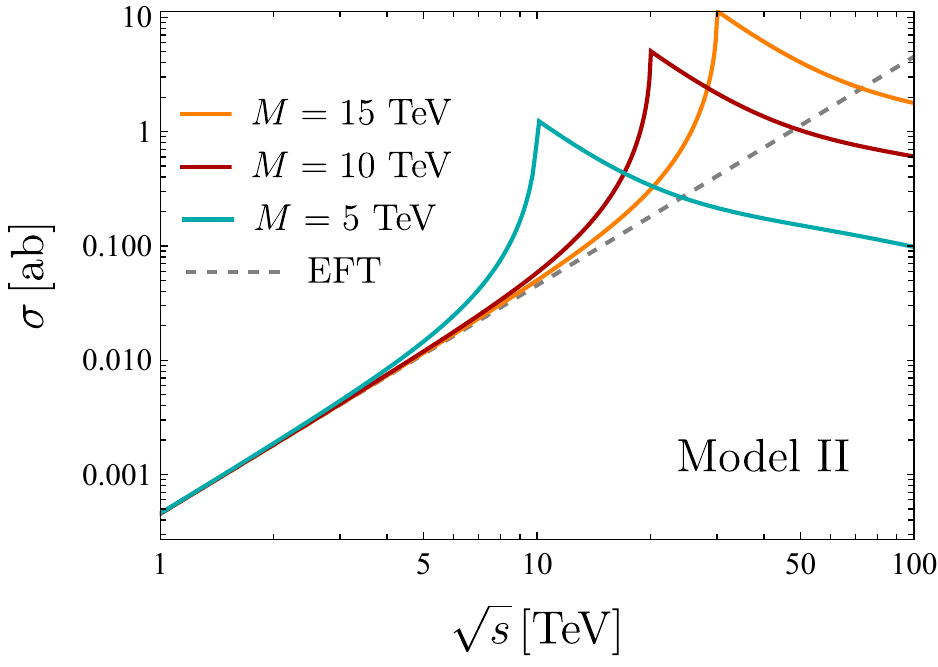}
	\caption{Left (right): Cross section of $\mu^+ \mu^- \rightarrow h \gamma$ for $M=(5,10,15)$ TeV and $X=1/2$ for the simplified model I (II)
	in the representation $R = 121$ (see Appendix A). The NP couplings have been fixed to solve the $(g-2)\mul$ discrepancy. 
	The dashed lines correspond to the EFT prediction.}
	\label{fig:s3sigma}	
\end{figure}

The EFT contribution to the $\mu^+\mu^-\to h \gamma $ amplitude is encapsulated in the form-factors $\mathcal{D}^{(\prime)}$,
\begin{align}
\label{eq:ff-eft-D}
\mathcal{D}_\mathrm{EFT} = -i{\sqrt{2}\, \mathrm{Re}({C_{e\gamma}})}\,,\qquad\qquad \mathcal{D}^\prime_\mathrm{EFT} = {\sqrt{2}\, \mathrm{Im}({C_{e\gamma}})}\,,
\end{align}
whereas $\mathcal{F}^{(\prime)}=0$ at this order in the EFT expansion. By using the effective coefficients $C_{e\gamma}$ given in Eq.~\eqref{eq:Deltaamu_I}--\eqref{eq:Deltaamu_II} for the simplified models I and II, respectively, we retrieve the first term in the $s/M$ power expansion of $\mathcal{D}^{(\prime)}$ in Sec.~\ref{ssec:full}, which is an important cross-check of our results.
After integrating over $t$, the total $\mu^+\mu^-\to h\gamma $ cross section reads
\begin{align}
\label{eq:sigma-eft}
\sigma_{h\gamma}^\mathrm{EFT} =  \dfrac{s\,  |C_{e\gamma}|^2}{48\pi} \approx 0.7~\mathrm{ab} \left(\dfrac{\sqrt{s}}{30~\mathrm{TeV}}\right)^2 \left(\dfrac{\Delta a_\mu}{3\times 10^{-9}}\right)^2\,.
\end{align}

\noindent
in agreement with Ref. \cite{Buttazzo:2020ibd}. It is clear from this equation that the sensitivity on $\Delta a_\mu$ increases with $\sqrt{s}$, as long as the EFT approach is valid. The energy scale $\sqrt{s}$ at which the total cross section departs from the EFT predictions for a given mass $M$ of the NP states will be derived along with our numerical results in the following section.

In principle, the operator $\mathcal{O}_{eH}=(H^\dagger H) \, \bar{\ell} e H$ would also contribute to the $\mu^+\mu^-\to  h \gamma$ process via a modification of the Higgs couplings to muons. However, this effect scales as $\sigma \propto 1/s$ and, therefore, it is safely negligible at high-energies compared to the NP contributions discussed above.
Interestingly, $\mathcal{O}_{eH}$ also contributes to the processes $\mu^+\mu^-\to h h$ and $\mu^+\mu^- \to h h h$~\cite{Dermisek:2021mhi}. 
In particular, we find that the cross section of $\mu^+\mu^-\to h h$ is independent of the collider energy, whereas the one of $\mu^+\mu^- \to h h h$ grows linearly with $s$ and it has a comparable 
size to the $\mu^+\mu^-\to  h\gamma$ cross section.
However, in order to determine the NP sensitivity of $\mu^+\mu^-\to h h (h)$, a detailed study of the SM background~\cite{Chiesa:2020awd} is mandatory,
which is beyond the scope of the present paper.~\footnote{Note, also, that the dependence of the Wilson coefficient of $\mathcal{O}_{eH}$ on the NP couplings differs from the one of $C_{e\gamma}$ by a factor of $(A/M)^2$ (model I) or $\kappa^2$ (model II), thus not being in direct correspondence with $\Delta a\mul$.} 

\begin{figure}[!ht]
		\includegraphics[width = 0.49\textwidth]{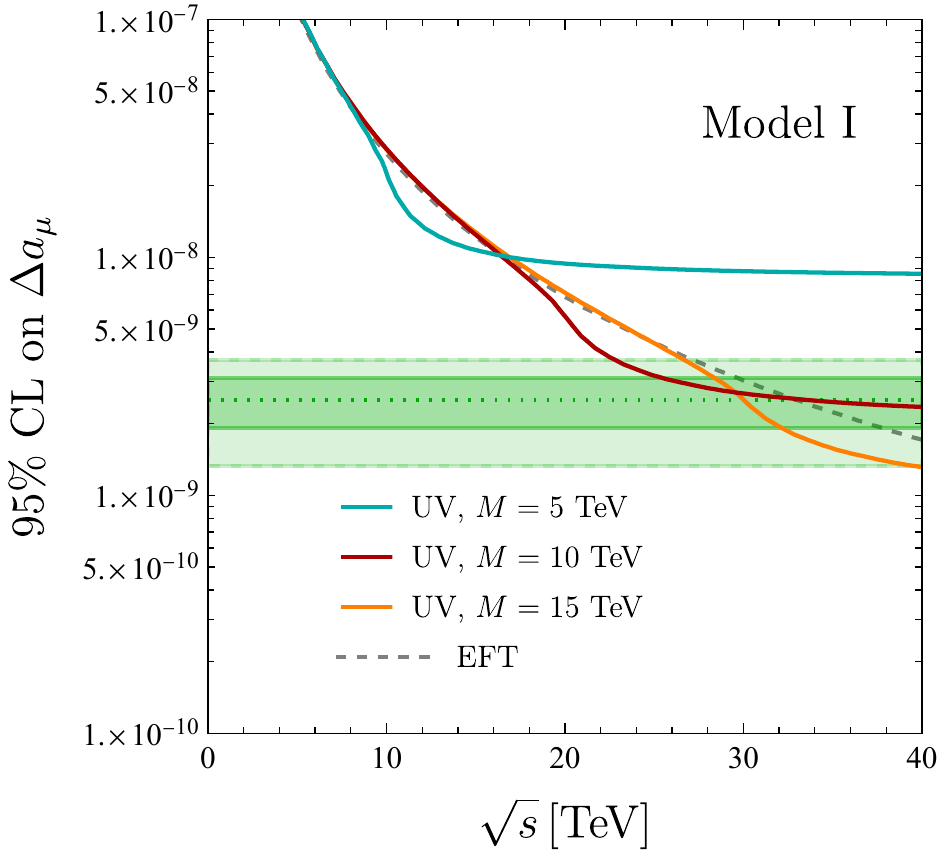}\;\;
		\includegraphics[width = 0.49\textwidth]{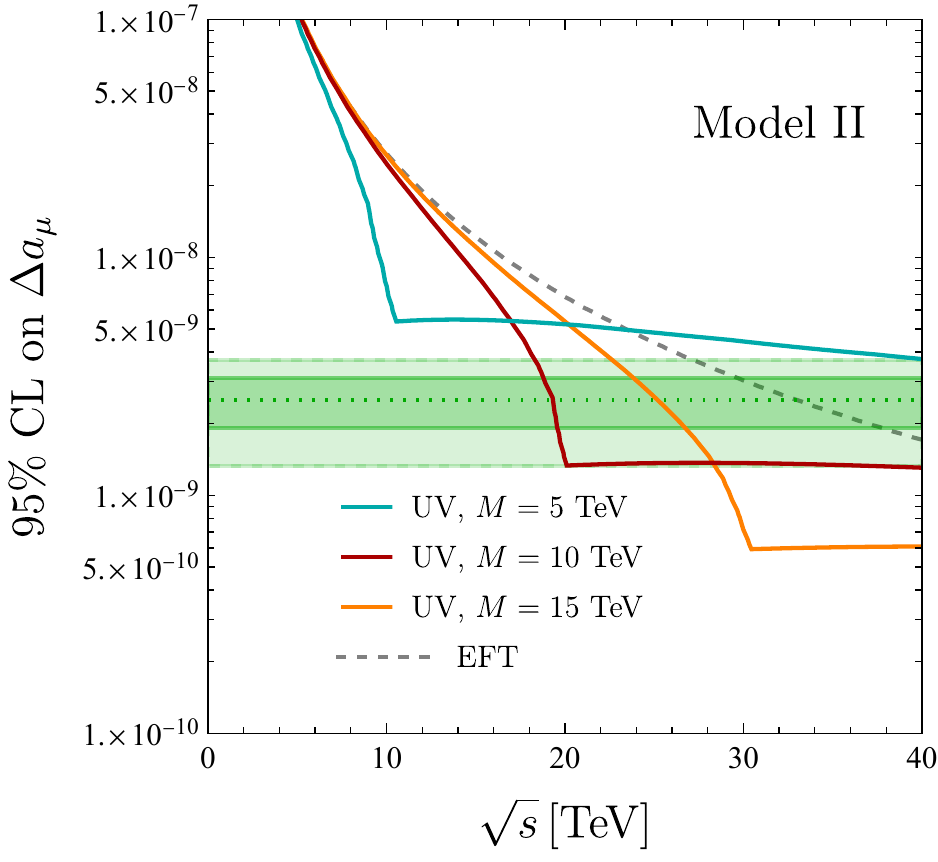}
	\caption{95\% C.L.\ reach on $\Delta a_\mu$ as a function of the center-of-mass energy $\sqrt{s}$ from the process 
	$\mu^+\mu^-\to h\gamma$ in EFT (dashed line) and in the simplified models I (left) and II (right) for three reference NP masses $M=(5,10,15)$ TeV and for the hypercharge $X=1/2$. The  darker (lighter) green bands represent the $1\sigma$ ($2\sigma$) ranges for $\Delta a\mul$ given by Ref.~\cite{Aoyama:2020ynm}.}
	\label{fig:hgamma}
\end{figure}

\subsection{Numerical results}
\label{ssec:numerical}

First, we report in Fig.~\ref{fig:s3sigma} the total cross section of $\mu^+ \mu^- \rightarrow h \gamma$ as a function of $\sqrt{s}$ for the simplified models I (left panel) and II (right panel), with the representation $R=121$ and hyperchage $X=1/2$ taken as our benchmark. The masses of the new states are fixed to three reference values, namely $M=(5,10,15)$ TeV, and the NP couplings are fixed to solve the $(g-2)\mul$ discrepancy.
Interestingly, we confirm numerically that the EFT agrees with the UV theory remarkably well for energies as large as $\sqrt s \sim M$ in both scenarios. 
At $\sqrt s \sim 2M$, the UV cross section shows a resonance peak corresponding to the fact that the virtual particles can then be produced on-shell.
Finally, for energies $\sqrt s \gtrsim 2M$ the UV cross section scales as $1/s$, as expected by the unitarity of the $S$-matrix. The most prominent difference between the two models is that in the model II the resonance is larger and more peaked than in the model I.

Next, following the analysis of Ref.~\cite{Buttazzo:2020ibd}, we study the capability of the process $\mu^+\mu^- \to h \gamma$ to probe the muon $g$-2 anomaly.
The SM irreducible $\mu^+\mu^- \to h\gamma$ background is small and can be neglected for $\sqrt{s} \gg 1$~TeV.
Instead, the main source of background comes from the $\mu^+\mu^- \to Z\gamma$ process, where the $Z$ boson is misreconstructed as a Higgs boson. 
An efficient way to isolate the $h\gamma$ signal from the background is to exploit the different angular distributions of the two processes, requiring that 
the probability of misreconstructing a $Z$ boson as a Higgs is less than 10\%.

In Fig.~\ref{fig:hgamma}, we show the 95\% C.L.\ reach from $\mu^+\mu^-\to h\gamma$ on the anomalous magnetic moment of the muon as a function of the center-of-mass energy $\sqrt{s}$ for the simplified model I (left panel) and II (right panel), and for three reference NP masses $M=(5,10,15)$ TeV. The dashed lines correspond to the EFT limit derived from Eq.~\eqref{eq:sigma-eft}. The NP sensitivity has been obtained imposing 
that the significance of the signal satisfy $S>2$, where $S=N_S/\sqrt{N_B+N_S}$, and $N_{S}$ and $N_B$ denote the number of signal and background events, respectively. The number of events is estimated considering a $\bar b b$ final state, with an $80 \%$ $b-$tagging efficiency and by imposing the kinematical cut $\left|\cos \theta\right|<0.6$ for which $S$ is maximized, where $\theta$ denotes the photon scattering-angle.

As already discussed above, the EFT result is accurately reproduced for $\sqrt s \lesssim M$. On the other hand, for energies close to the resonant-production threshold, i.e.~$\sqrt s = 2M$,
the simplified models have an even higher sensitivity to $\Delta a\mul$ than the EFT, especially in the case of the model II.
Finally, for energies $\sqrt s \gg 2M$, the $\mu^+\mu^-\to h\gamma$ cross section in the simplified models scales as $\sigma \sim 1/s$ and therefore 
the number of signal events $N_S \propto \sigma \times \mathcal{L}$ becomes constant with respect to the energy, since the luminosity scales as $\mathcal{L}\propto s$~\cite{Delahaye:2019omf}. 
This behavior is in contrast with the EFT expectation, for which $\sigma \sim s$ and therefore $N_S \propto s^2$.

Although our simplified models can account for the muon g-2 anomaly only for $\sqrt s\gtrsim M$, where the EFT description breaks down, the capability of the process $\mu^+ \mu^- \rightarrow h \gamma$ to probe NP effects in $\Delta a \mul$ is confirmed provided that the mass $M$ is sufficiently large, as shown in Fig. \ref{fig:hgamma}. On the other hand, for light mediators, the $\mu^+\mu^-\to h\gamma$ process is no longer able to probe NP effects in $\Delta a\mul$. In this case, it is more convenient to directly produce the new states instead of probing them indirectly, as we explore in the following.

\section{Direct high-energy probes of the muon \boldmath{$g-2$}}
\label{section:2_to_3}

In this section, we will analyse the capability of a high-energy muon collider to discover the new particles of the simplified models I and II via their direct production.
In particular, since we assume an underlying $Z_2$ symmetry, the new states are always produced in pairs.
Since the muon $g$-2 anomaly can be typically accommodated for $M \lesssim 15$ TeV, it follows that a high-energy muon collider running with energies 
$\sqrt{s}\gtrsim 30$ TeV should be able to directly observe these new particles. 

\subsection{\boldmath{$2\to 2$} processes}
\label{sec:2to2}

\begin{figure}[!htb]
	\centering
	\includegraphics[width=0.6\textwidth]{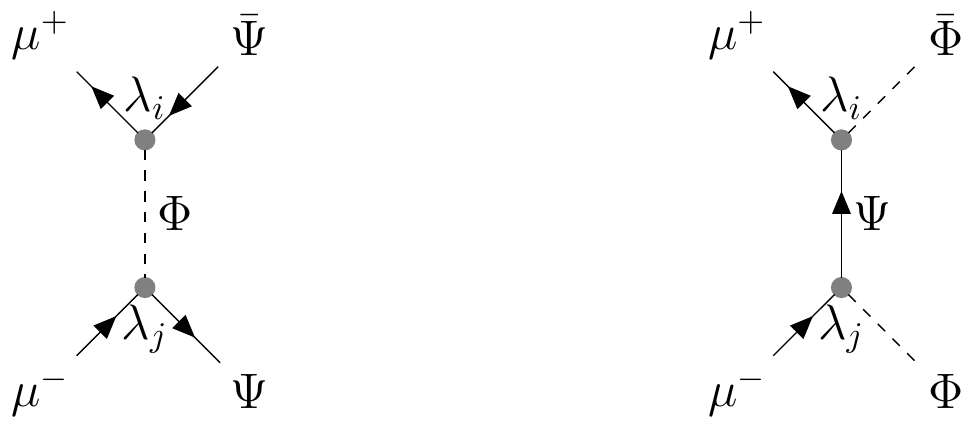}
	\caption{\label{fig:s5diag22}Feynman diagrams for $2\rightarrow2$ pair production processes in the simplified models I and II.}
\end{figure}

The most relevant $2\rightarrow2$ processes are schematically represented by the Feynman diagrams of Fig.~\ref{fig:s5diag22}, 
where $\Phi, \Psi$ refer collectively to the scalar and fermion fields of the models I or II. In the following, we report the analytical expressions of the relevant cross sections. 

For the model I, we obtain
\begin{align}
	\sigma ( \mu ^+ \mu^- \rightarrow \bar \Phi_i \Phi_i) &= \frac{\eta\left| \lambda_i ^{\text{I}}\right|^4}{32 \pi s} \left[\tanh \inv \left(\sqrt{1-\frac{4M^2}{s}}\right) - \sqrt{1-\frac{4M^2}{s}}\right]
	\,, \nonumber \\[0.2em]
	\label{eq:2_to_2_I}
	\sigma ( \mu ^+ \mu^- \rightarrow \bar \Phi_i \Phi_j) &= \frac{\eta \left| \lambda_i ^{\text{I}} \lambda _j ^{\text{I}}\right|^2}{64 \pi s} \sqrt{1-\frac{4M^2}{s}} 
	\,,\\[0.2em]
	\sigma ( \mu^+ \mu^- \rightarrow \bar \Psi \Psi) &= \frac{\left| \lambda_E ^{\text{I}}\right|^4 +\eta\left|\lambda _L ^{\text{I}}\right|^4}{64 \pi s} \sqrt{1-\frac{4M^2}{s}}  
	\,,  \nonumber
\end{align}
whereas for the model II
\begin{align}
	\begin{split}
		\label{eq:2_to_2_II}
		\sigma (\mu ^+ \mu^-\rightarrow \bar \Psi_i \Psi_i) &= \frac{\eta \left| \lambda_i ^{\text{II}}\right|^4 }{64 \pi s} \sqrt{1-\frac{4M^2}{s}} 
		\,, \\[0.2em]
		\sigma (\mu ^+ \mu^- \rightarrow \bar \Psi_i \Psi_j) &= \frac{\eta \left| \lambda_i ^{\text{II}} \lambda _j ^{\text{II}}\right|^2}{64 \pi s} \sqrt{1-\frac{4M^2}{s}} 
		\,,  \\[0.2em]
		\sigma (\mu ^+ \mu^- \rightarrow \bar \Phi \Phi) &= \frac{\left| \lambda_E ^{\text{II}}\right|^4 +\eta \left|\lambda _L ^{\text{II}}\right|^4}{32 \pi s} \left[\tanh \inv \left(\sqrt{1-\frac{4M^2}{s}}\right) - \sqrt{1-\frac{4M^2}{s}}\right]
		,
	\end{split}
\end{align}
where $i,j \in \lbrace E, L \rbrace$, $i\neq j$
and the factor $\eta$ is equal to $1$ for the representations $R=\{121,212\}$, while $\eta=1$ or $4$ for $R=\{323,232\}$ depending on the specific final state~\footnote{In the presence of an $SU(2)_L$ triplet $\chi^a$ ($a=1,2,3$), the electric charge eigenstates are $\chi_{\pm}=\frac{\chi^1 \mp i \chi^2}{\sqrt 2}$ and $\chi_3$.
Then, in the model I, $\eta=4$ for the final states
$(\bar \Psi_- \Psi_ - ), (\bar \Phi_{L,\frac{1}{2}} \Phi_{L,\frac{1}{2}}), (\bar \Phi_{L,\frac{1}{2}} \Phi_{E,+}), ( \Phi_{L,\frac{1}{2}} \bar \Phi_{E,+})$ 
of the representation \text{$R=323$}, and $(\bar \Psi_{\frac{1}{2}} \Psi_ {\frac{1}{2}}), (\bar \Phi_{L,-} \Phi_{L,-}), (\bar \Phi_{L,-} \Phi_{E,-\frac{1}{2}} ), ( \Phi_{L,-} \bar \Phi_{E,-\frac{1}{2}} )$ for \text{$R=232$}. In all other cases $\eta=1$. The lower index $\pm \frac{1}{2}$ refers to the component of the isospin doublets. For model II the situation is completely analogous.
}.
For the references values $\sqrt{s} = 30$ TeV and $M = 10$ TeV, the above cross sections attain comparable values of order $10^4\, \text{ab}$.
Note that in Eqs. (\ref{eq:2_to_2_I}) and (\ref{eq:2_to_2_II}), we have neglected the contributions stemming from the s-channel exchange of the SM gauge bosons 
$\gamma$ and $Z$. Indeed, since the solution of the muon $g$-2 anomaly requires $\lambda_{E,L} \gtrsim 2$, the t-channel diagrams of Fig.~\ref{fig:s5diag22} 
are by far dominant.

\subsection{\boldmath{$2\to 3$} processes}

\begin{figure}[!htb]
	\centering
	\includegraphics[width=0.65\textwidth]{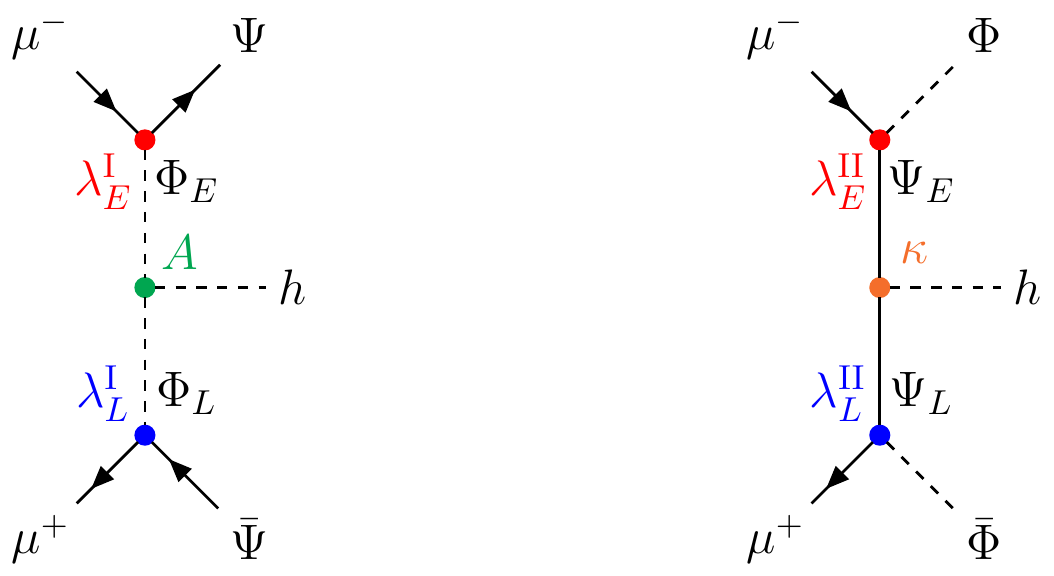}
	\caption{\label{fig:Feynman_III}Feynman diagrams contributing to the direct production channels $ \mu^+ \mu^- \rightarrow h \bar \Psi \Psi, h \bar \Phi \Phi$
	 in the simplified models I (left) and II (right).}
\end{figure}

Although the $2 \to 2$ processes shown in Fig.~\ref{fig:s5diag22} are unavoidably induced once a NP contribution to the muon $g$-2 is generated, it is important to stress that their observation cannot be promoted by any means as an unambiguous test of the $\Delta a_\mu$ anomaly.
Indeed, $\Delta a_\mu$ and the cross sections in Eq.~\eqref{eq:2_to_2_I} and \eqref{eq:2_to_2_II} depend on different combinations of the simplified models parameters. Therefore, it would be desirable to identify high-energy processes (if any) that are in one-to-one correspondence with the NP effects entering the muon $g$-2.
Interestingly, such processes do exist and are given by $\mu ^+\mu ^- \rightarrow h\bar \Psi \Psi $ in model I and by $ \mu ^+\mu ^-\rightarrow h \bar \Phi \Phi $ in model II, 
as illustrated in Fig.~\ref{fig:Feynman_III}.~\footnote{Other possibilities are given by $\mu ^+\mu ^- \rightarrow h\bar \Phi_{E(L)} \Phi_{E(L)} $ in model I and
$\mu ^+\mu ^- \rightarrow h\bar \Psi_{E(L)} \Psi_{E(L)} $ in model II, where the Higgs is emitted from the final state legs. Instead, the analogous processes $\mu ^+\mu ^- \rightarrow h \bar \Phi_{E(L)} \Phi_{L(E)} $ and $\mu ^+\mu ^- \rightarrow h  \bar \Psi_{E(L)} \Psi_{L(E)} $ do not depend on the same combination of NP couplings as $\Delta a_\mu$.
A careful analysis of the different final state products and of the Higgs kinematical properties may be exploited to disentangle these processes.}

This correspondence is evident by comparing this diagram with the one for $\Delta a_\mu$ depicted in Fig.~\ref{fig:Feynman_I}. In particular, the correlation between these processes with $\Delta a_\mu$ is exact for a degenerate spectrum of the new states, which we consider in the following. For a general spectrum, this may not be necessarily the case and a dedicated analysis would be required.

The differential cross sections for these processes can be written in the following form,
\begin{align}
	d\sigma_{2\rightarrow 3} = 
	\frac{|\mathcal{\overline A}|^2}{256 \pi^3 } \frac{\lambda^{1/2}(0,s,q^2) \lambda^{1/2}(q^2,M^2,M^2)}{s^2 q^2} 
	\frac{ d\cos\theta_1}{2} \frac{d\phi_1}{2\pi}\frac{ d\cos\theta_2 }{2}\frac{d\phi_2}{2\pi} dq^2 
\end{align}
where $\lambda(x,y,z)=x^2+y^2+z^2- 2(xy+xz+yz)$ is the K\"{a}ll\'{e}n function, and we define $q = k_1 + k_2$, with $q^2 \in [4M^2,s]$. The angles $\theta_i$ and $\phi_i$ are defined in Fig.~\ref{fig:appBframes} and are integrated in the ranges $\theta_i \in [0,\pi]$ and $\phi_i \in [0,2\pi]$. For a detailed discussion about the non-trivial kinematics of $2\to 3$ processes we refer to appendix \ref{app:kinematics}. The $|\mathcal{\overline A}|^2$ expression for the process $\mu ^+(p_b)  \mu ^-(p_a ) \rightarrow h(k) \bar \Psi(k_2) \Psi(k_1) $ in the model I reads\footnote{In this case, $\eta=4$ for $\mu^+ \mu^- \rightarrow h\bar \Psi_- \Psi_ -   \;(R=323)$ and $\mu^+ \mu^- \rightarrow h \bar \Psi_{\frac{1}{2}} \Psi_{\frac{1}{2}}  \; (R=232)$, while $\eta=1$ otherwise.}
\begin{align}
	|\mathcal{\overline A}|^2 = \frac{\eta}{4}\frac{|\lambda_L ^{\text{I}} \lambda_E ^{\text{I}} A|^2}{[M^2 - (p_a-k_1)^2][M^2- (p_b-k_2)^2]}\,.
	\label{eq:s4ampI}
\end{align}
Similarly, the squared amplitude $ \mu ^+(p_b)  \mu ^-(p_a )  \rightarrow h(k) \bar \Phi(k_2) \Phi(k_1)$ in the model II is given by
\begin{flalign}
	|\mathcal{\overline A}|^2 =  \frac{2 \eta |\lambda_L ^{\text{II}}\lambda_E ^{\text{II}} \kappa|^2 M^2}{[M^2 - (p_a-k_1)^2]^2 [M^2- (p_b-k_2)^2]^2}  
	\bigg[  p_a \cdot (k_2-k_1) \,  p_b \cdot(k_1-k_2)
	\nonumber\\
	\qquad-2 \left(p_a \cdot k_1\, p_b \cdot k_2 -\frac{sM^2}{4}\right) \left(1-\frac{k_1 \cdot k_2 }{2M^2}\right)
	\bigg]\,,
	\label{eq:s4ampII}
\end{flalign}
where the scalar products can be easily computed by using the expressions from appendix~\ref{app:kinematics}. 

\begin{figure}[!ht]
	\centering
	\hfill \begin{subfigure}[t]{0.49\textwidth}
		\includegraphics[width=\textwidth]{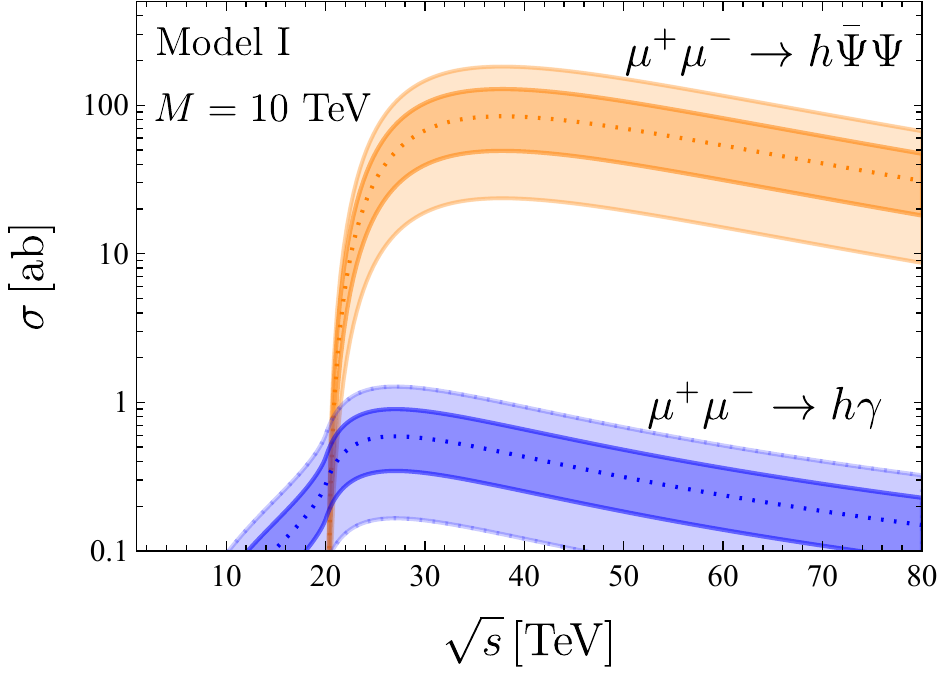}
	\end{subfigure}
	\begin{subfigure}[t]{0.49\textwidth}
		\includegraphics[width=\textwidth]{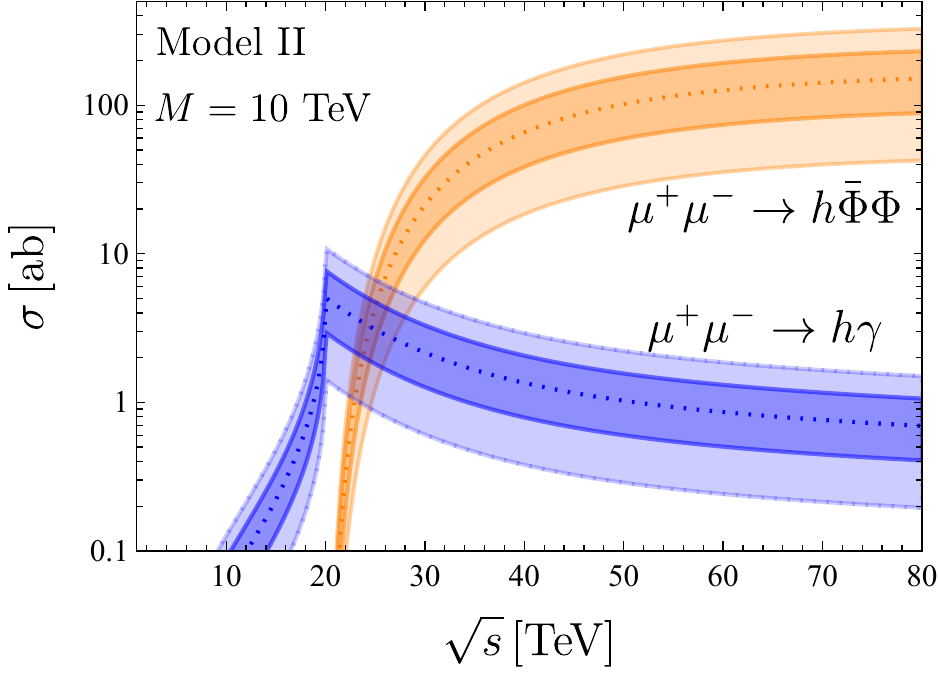}
	\end{subfigure}
\hfill
\caption{\label{fig:s5sig23}
Cross sections of $\mu^+\mu^-\to h\gamma$ and $\mu^+\mu^- \rightarrow h \bar \Psi \Psi, h\bar \Phi \Phi$ vs. $\sqrt{s}$ setting $X=1/2$ and $M=10$ TeV 
in the simplified model I (left) and II (right). Inner (outer) regions correspond to $\Delta a \mul$ values in the $1\sigma$ ($2\sigma$) allowed range.}
\end{figure}

In Fig.~\ref{fig:s5sig23}, we report the cross sections of $\mu^+\mu^- \rightarrow  h\bar \Psi \Psi , h\bar \Phi \Phi $ vs.~$\sqrt{s}$ setting $X=1/2$ and $M=10$ TeV. 
Inner (outer) regions correspond to $\Delta a \mul$ values in the $1\sigma$ ($2\sigma$) allowed range.  As expected, the $2\to 3$ cross sections are typically 2-3 orders of magnitude smaller than those for $2\rightarrow2$ processes (see Eqs.~(\ref{eq:2_to_2_I}) and (\ref{eq:2_to_2_II})) 
due to the additional phase-space suppression. However, at a muon collider running at $\sqrt s = 30$ TeV, we expect several hundreds events for these $2\to 3$ processes that can 
be discriminated from the background thanks to the coincidence of two NP particles accompanied by a Higgs boson.
In Fig.~\ref{fig:s5sig23}, we also plot the cross section of $\mu^+\mu^-\rightarrow h \gamma$ as a function of $\sqrt{s}$ to stress its interplay with the direct search processes 
$\mu^+\mu^- \rightarrow h \bar \Psi \Psi$ and $\mu^+\mu^- \rightarrow h\bar \Phi \Phi$. Indeed, for collider energies below the threshold required for a 
direct production of new states, $\mu^+\mu^- \rightarrow h \gamma$ provides a unique way to test the muon $g$-2 anomaly. Instead, when the final states
$h \bar \Psi \Psi$ and $h \bar \Phi \Phi$ are kinematically allowed, the study of their correlation with the process $\mu^+\mu^- \rightarrow h \gamma$
would still be of great importance to pin down the details of the underlying NP model.

\section{Conclusions}

In this paper, we have studied the solution of the muon $g$-2 anomaly through new physics scenarios with heavy scalars and vectorlike fermions appearing above the TeV-scale. 
Such a solution is only viable provided that a chiral enhancement is at work, which in turn requires that the new states couple to the SM Higgs boson.
As already emphasised in the EFT context, a muon collider running at center-of-mass energies $\sqrt{s}$ in the multi-TeV range 
would be the ideal machine to test this anomaly model-independently through the study of the $\mu^+\mu^-\to h\gamma$ process~\cite{Buttazzo:2020ibd}. 

We have explored the connection between $\Delta a_\mu$ and $\mu^+\mu^-\to h\gamma$ in the context of the concrete NP scenarios mentioned above, which contribute to both 
observables at one-loop level, extending the EFT results to the case where $\sqrt{s}$ is larger than the mass of the new particles. In particular, we have found that the EFT 
approach describes remarkably well the $\mu^+\mu^-\to h\gamma$ cross section for $\sqrt{s}$ values even at the vicinity of the EFT cutoff, where the EFT description 
is expected to break down, as shown in Figs.~\ref{fig:s3sigma} and \ref{fig:hgamma}. These results confirm and reinforce the complementarity of $\Delta a_\mu$ with the high-energy process 
$\mu^+\mu^-\to h\gamma$.

Another goal of this work has been to study the direct search signatures of our simplified models, as well as their interplay with the indirect search $\mu^+\mu^-\to h\gamma$. 
If kinematically allowed, the processes $\mu^+ \mu^- \rightarrow \bar \Psi \Psi, \bar \Phi \Phi$, where $\Psi$ and $\Phi$ refer to heavy vectorlike fermions and scalars, 
are unavoidably induced with sizable cross sections.
However, the cross sections of these $2\to 2$ processes are not directly correlated with $\Delta a_\mu$ as they depend on different combinations of NP couplings. 

We have shown in this paper that the cross sections of the processes $\mu^+\mu^- \rightarrow h\bar\Psi \Psi, h\bar \Phi \Phi$ with a Higgs boson in the final state, 
which we have computed under the assumption of degenerate masses, are in one-to-one correspondence with the NP effects entering the muon $g$-2, as shown in Fig.~\ref{fig:Feynman_I} and~\ref{fig:Feynman_III}.
Although suppressed by two orders of magnitude compared to the $2\rightarrow2$ processes, 
due to the additional phase-space suppression, we still expect several hundreds of events for $\mu^+\mu^-  \rightarrow h \bar \Psi \Psi, h\bar \Phi \Phi$ at a muon collider running at 
$\sqrt s = 30$ TeV, which can be discriminated thanks to the coincidence of two NP particles together with a Higgs boson in the final state.
As shown in Fig.~\ref{fig:s5sig23}, there is an interesting interplay between the indirect probe $\mu^+\mu^- \rightarrow h \gamma$ and the direct ones
$\mu^+\mu^- \rightarrow h \bar \Psi \Psi$ and $\mu^+\mu^- \rightarrow h\bar \Phi \Phi$. Indeed, for collider energies below the threshold required for a direct production of new states, $\mu^+\mu^- \rightarrow h \gamma$ provides a unique way to access the muon $g$-2 anomaly. Instead, when $\mu^+\mu^- \rightarrow h \bar \Psi \Psi$ and $\mu^+\mu^- \rightarrow h \bar \Phi \Phi$ are kinematically allowed, they are typically the best probe of $\Delta a_\mu$. This complementarity illustrates the fact that a correlated study of direct and indirect new physics signals at a muon collider would be a powerful handle to disentangle among the underlying model accommodating 
the $\Delta a_\mu$ anomaly.

\section{Acknowledgments}
\label{sec:acknowledgments}
We thank A. Wulzer for useful discussions.
This project has received support from the European Union’s Horizon 2020 research and innovation programme under the Marie Skłodowska-Curie grant agreement No~860881-HIDDeN and by the INFN Iniziativa Specifica APINE.

\clearpage

\appendix
\section{Simplified model Lagrangians and RGEs}
\label{app:running}
In this appendix, we collect the relevant expressions for the running of the fundamental parameters of the simplified models I and II introduced in section I.
Assuming a $Z_2$ symmetry to avoid mixing of the new states with SM fields and allowing $SU(2)_L$ representations up to triplets, the most generic Lagrangians are~\cite{Crivellin:2021rbq}:
\begin{align}
	\Lagr_\text{I}^{121} &= \lambda _L^\text{I}\,\bar\ell^a\Psi\Phi _L^a + \lambda _E^\text{I}\,\bar e\Psi\Phi _E + A\,\Phi _L^{a\dag }\Phi _E{H^a}+\mathrm{h.c.}\,,\notag\\[0.3em]
	\Lagr_\text{II}^{121} &= \lambda _L^\text{II}\,\bar\ell^a\Phi\Psi _L^a + \lambda _E^\text{II}\,\bar e\Psi _E{\Phi } + \kappa\, \bar \Psi _L^a\Psi _E{H^a}+\mathrm{h.c.}\,\notag\\[0.3em]
	\Lagr_\text{I}^{212} &= \lambda _L^\text{I}\,\bar\ell^a\Phi _L{\Psi ^a} + \lambda _E^\text{I}\,\bar e\Psi _{}^a(i\tau_2\Phi _E)^a + A\,{(i\tau_2 H)^a} \Phi _L^\dag \Phi _E^{a}+\mathrm{h.c.}\,\notag\\[0.3em]
	\Lagr_\text{II}^{212} &= \lambda _L^\text{II}\,\bar\ell^a\Psi _L{\Phi ^a} + \lambda _E^\text{II}\,\bar e(i\tau_2\Psi _E)^a{\Phi^a} + \kappa\, \bar \Psi_L{\left( {i{\tau _2}H} \right)^a}\Psi _E^a+\mathrm{h.c.}\,\notag\\[0.3em]
	\Lagr_\text{I}^{323} &= \lambda _L^\text{I}\,\bar\ell^a{\left( {\tau \cdot\Psi } \right)_{ab}}\Phi _L^b + \lambda _E^\text{I}\,\bar e\Psi _{}^\alpha\Phi _E^\alpha + A\,\Phi _L^{a\dag }{\left( {\tau \cdot\Phi _E^{}} \right)_{ab}}{H^b}+\mathrm{h.c.}\,\notag\\[0.3em]
	\Lagr_\text{II}^{323} &= \lambda _L^\text{II}\,\bar\ell^a{\left( {\tau \cdot\Phi } \right)_{ab}}\Psi _L^b + \lambda _E^\text{II}\,\bar e\Psi _E^\alpha{\Phi ^\alpha} + \kappa\, \bar \Psi _L^a{\left( {\tau \cdot{\Psi _E}} \right)_{ab}}{H^b}+\mathrm{h.c.}\,\notag\\[0.3em]
	\Lagr_\text{I}^{232} &= \lambda _L^\text{I}\,\bar\ell^a{\left( {\tau \cdot{\Phi _L}} \right)_{ab}}{\Psi ^b} + \lambda _E^\text{I}\,\bar e\Psi _{}^a(i\tau_2\Phi _E)^a + A\,(i\tau_2 H)^a(\tau\cdot \Phi _L^\dagger)_{ab}\Phi _E^{b}+\mathrm{h.c.}\,\notag\\[0.3em]
	\Lagr_\text{II}^{232} &= \lambda _L^\text{II}\,\bar\ell^a{\left( {\tau \cdot\Psi _L^{}} \right)_{ab}}{\Phi ^b} + \lambda _E^\text{II}\,\bar e(i\tau_2\Psi _E)^a{\Phi^a} + \kappa\, {\left( {\tau \cdot{{\bar \Psi }_L}} \right)_{ab}}{\left( {i{\tau _2}H} \right)^a}\Psi _E^b+\mathrm{h.c.}\,
	\label{eq:s1lagrangians}
\end{align}
where $a,b$ denote $SU(2)_L$ indices, $\tau$ are the Pauli matrices and the charges of the various fields are defined in Table~\ref{tab:s1charges}.

The RGEs for the simplified model Yukawa-couplings and for the SM gauge couplings, as well as for the Higgs boson quartic coupling in the models of type-II \footnote{In the models of type-I, the Higgs quartic does not receive BSM contributions at one-loop.}, have been calculated using the tool \texttt{RGBeta} \cite{Thomsen:2021ncy}. They read
\begin{align}
	\text{Model I:  }
	\begin{cases}
		(4\pi)^2 \frac{d g^2}{d\log \mu} =  \beta _g g^4 \\
		(4\pi)^2 \frac{d g^{\prime 2}}{d\log \mu} = \beta _{g'} g^{\prime 4} \\
		(4\pi)^2 \frac{d\lambda_L ^{\text{I}}}{d\log \mu} = \lambda_L ^{\text{I}} \left[-\beta _L ^g g^2 -\beta _L ^{g'}g^{\prime 2}+ \beta _L ^L \lambda_L ^{\text{I}\,2} + \beta _L ^y y\mul ^2   \right]\\
		(4\pi)^2 \frac{d\lambda_E ^{\text{I}} }{d\log \mu} = \lambda_E ^{\text{I}} \left[-\beta _E ^g g^2 -\beta _E ^{g'}g^{\prime 2}+ \beta _E ^E \lambda_E ^{\text{I}\,2}+ \beta _E ^y y\mul ^2   \right]\\
	\end{cases},
	\label{eq:s4RGEIgen}
\end{align}
\begin{align}
	\text{Model II:  }
	\begin{cases}
		(4\pi)^2 \frac{d g^2}{d\log \mu} =  \beta _g g^4 \\
		(4\pi)^2 \frac{d g^{\prime 2}}{d\log \mu} = \beta _{g'} g^{\prime 4} \\
		(4\pi)^2 \frac{d\lambda_L ^{\text{II}}}{d\log \mu} = \lambda_L ^{\text{II}} \left[-\beta _L ^g g^2 -\beta _L ^{g'}g^{\prime 2}+ \beta _L ^L \lambda_L ^{\text{II}\, 2} +  \beta _L ^E \lambda_E ^{\text{II}\, 2}+ \beta _L ^{\kappa} \kappa ^2 + \beta _L ^y y\mul ^2   \right] + \beta_L ^{yE\kappa} y\mul \lambda_E ^{\text{II}} \kappa\\
		(4\pi)^2 \frac{d\lambda_E ^{\text{II}}}{d\log \mu} = \lambda_E ^{\text{II}} \left[-\beta _E ^g g^2 -\beta _E ^{g'}g^{\prime 2}+ \beta _E ^L \lambda_L ^{\text{II}\, 2} + \beta _E ^E \lambda_E ^{\text{II}\, 2} + \beta _E ^{\kappa} \kappa ^2 +\beta _E  ^y y\mul ^2   \right] + \beta _E ^{yL\kappa}  y\mul \lambda _L ^{\text{II}}\kappa \\
		(4\pi)^2 \frac{d\kappa}{d\log \mu} = \kappa \left[-\beta _{\kappa} ^g g^2 -\beta _{\kappa} ^{g'}g^{\prime 2}+ \beta _{\kappa} ^L \lambda_L ^{\text{II}\, 2} + \beta _{\kappa} ^E \lambda_E ^{\text{II}\, 2} + \beta _{\kappa} ^{\kappa} \kappa ^2+\beta _{\kappa}  ^y (y\mul ^2 +3 y_t^2)   \right] + \beta_{\kappa} ^{yLE} y\mul \lambda _L ^{\text{II}}\lambda_E ^{\text{II}} \\
		(4\pi)^2 \frac{d\lambda}{d\log \mu} = \beta _{\lambda} ^{\text{SM}} (g,g',\lambda, y_t)+\beta_{\lambda}^{\lambda \kappa} \lambda \kappa^2 - \beta_{\lambda}^{\kappa} \kappa ^4
	\end{cases}.
	\label{eq:s4RGEIIgen}
\end{align}
The values of the coefficients of the $\beta$-functions for the various representation are given in Table~\ref{table:beta_functions}, where they are written as vectors with components ordered as they appear in Eqs.~(\ref{eq:s4RGEIgen}) and (\ref{eq:s4RGEIIgen}).

\begin{table}[!htb]
	\centering
	{
		\renewcommand{\arraystretch}{1.3}
		\begin{tabular}{c|c|c}
			$R$ & Model I & Model II\\
			\toprule 
			\multirow{6}{*} {$121$} & $\beta_g = \beta_g ^{\text{SM}} + \frac{1}{3}$ &$\beta_g = \beta_g ^{\text{SM}} + \frac{4}{3}$ \\
			& $\beta_{g'}=  \beta_{g'} ^{\text{SM}} + \frac{1}{3}\left(\scriptstyle 3+8X+14X^2\right)$& $\beta_{g'}=  \beta_{g'} ^{\text{SM}} + \frac{2}{3}\left(\scriptstyle 6+16X+13X^2\right)$\\
			&$\beta_L ^i= \left\{ \frac{9}{4}, \frac{3}{4} ({\scriptstyle 1+4X^2 }), \frac{5}{2}, \frac{1}{2} \right\} $ & $\beta_L ^i = \left\{ \frac{9}{2}, \frac{3}{2} ({\scriptstyle 1+2X+2X^2 }), {\scriptstyle 3 }, {\scriptstyle 1}, \frac{1}{2}, \frac{1}{2}, {\scriptstyle 2} \right\} $ \\
			& $\beta_E ^i = \left\{ \scriptstyle 0, 3( 1+X^2), 2 ,1 \right\} $&
			$\beta_E ^i= \left\{ \scriptstyle 0, 3( 2+2X+X^2), 2 ,2, 1,1,4 \right\} $\\
			& &  $\beta_{\kappa} ^i= \left\{ \frac{9}{4}, \frac{15}{4}{\scriptstyle +9X+6X^2}, \frac{1}{4} ,\frac{1}{4}, \frac{7}{2}, {\scriptstyle 1}, {\scriptstyle 1} \right\} $\\
			& & $\beta_{\lambda} ^i= \left\{ \scriptstyle 4,2 \right\} $\\
			\midrule
			\multirow{6}{*} {$212$} & $\beta_g = \beta_g ^{\text{SM}} + \frac{5}{3}$& 
			$\beta_g = \beta_g ^{\text{SM}} + \frac{5}{3}$ \\
			& $\beta_{g'}=  \beta_{g'} ^{\text{SM}} + \frac{1}{6}\left(\scriptstyle 9+20X+44X^2\right)$&
			$\beta_{g'}=  \beta_{g'} ^{\text{SM}} + \frac{2}{3}\left(\scriptstyle 9+20X+14X^2\right)$\\
			& $\beta _L ^i = \left\{\frac{9}{2}, \frac{3}{4}({ \scriptstyle 1+4X^2 }), {\scriptstyle 3}, \frac{1}{2} \right\}$&
			$\beta_L ^i= \left\{ \frac{9}{4}, \frac{3}{2} ({\scriptstyle 1+2X+2X^2 }), {\frac{5}{2} }, {\scriptstyle 1},{\scriptstyle 1}, \frac{1}{2}, {\scriptstyle -2} \right\} $ \\
			& $\beta_E ^i= \left\{\frac{9}{4}, { \scriptstyle3(1+X^2) }, \frac{5}{2}, {\scriptstyle 1} \right\}$&
			$\beta_E ^i= \left\{ \frac{9}{4}, {\scriptstyle 3( 2+2X+X^2)}, {\scriptstyle 1} ,\frac{5}{2}, \frac{1}{2},{\scriptstyle 1},{\scriptstyle -2} \right\} $ \\
			& &  $\beta_{\kappa} ^i = \left\{ \frac{9}{4}, \frac{15}{4}{\scriptstyle +9X+6X^2}, \frac{1}{2} ,\frac{1}{4}, \frac{7}{2}, {\scriptstyle 1}, {\scriptstyle -1} \right\} $ \\
			& & $\beta_{\lambda} ^i= \left\{ \scriptstyle 4,2 \right\} $\\
			\midrule
			\multirow{6}{*} {$323$} & $\beta_g = \beta_g ^{\text{SM}} + {\scriptstyle 7}$&
			$\beta_g = \beta_g ^{\text{SM}} + {\scriptstyle 8}$ \\
			& $\beta_{g'}=  \beta_{g'} ^{\text{SM}} + \frac{1}{3}\left( \scriptstyle 7+16X+34X^2\right)$& 
			$\beta_{g'}=  \beta_{g'} ^{\text{SM}} + \frac{2}{3}\left(\scriptstyle 14+32X+23X^2\right)$\\
			& $\beta _L ^i = \left\{\frac{33}{4}, \frac{3}{4}({ \scriptstyle 1+4X^2} ), \frac{11}{2}, \frac{1}{2} \right\}$ &
			$\beta_L ^i= \left\{ \frac{9}{2}, \frac{3}{2} ({\scriptstyle 1+2X+2X^2 }), {\scriptstyle 5 }, {\scriptstyle 1},{\frac{3}{2}}, \frac{1}{2}, {\scriptstyle -2} \right\} $\\
			& $\beta_E ^i= \left\{ \scriptstyle 6,3(1+X^2),3, 1 \right\}$&
			$\beta_E ^i= \left\{ {\scriptstyle 6}, {\scriptstyle 3( 2+2X+X^2)}, {\scriptstyle 2} ,{\scriptstyle 3}, {\scriptstyle 1},{\scriptstyle 1},{\scriptstyle -4} \right\} $ \\
			& &  $\beta_{\kappa} ^i= \left\{ \frac{33}{4}, \frac{15}{4}{\scriptstyle +9X+6X^2}, \frac{3}{4} ,\frac{1}{4}, \frac{17}{2}, {\scriptstyle 1}, {\scriptstyle -1} \right\} $\\
			& & $\beta_{\lambda} ^i = \left\{ \scriptstyle 12,10 \right\} $\\
			\midrule
			\multirow{6}{*} {$232$} & $\beta_g = \beta_g ^{\text{SM}} + {\scriptstyle 3}$& 
			$\beta_g = \beta_g ^{\text{SM}} + {\scriptstyle 7}$ \\
			& $\beta_{g'}=  \beta_{g'} ^{\text{SM}} + \frac{1}{6}\left( \scriptstyle 11+28X+52X^2\right)$&
			$\beta_{g'}=  \beta_{g'} ^{\text{SM}} + \frac{2}{3}\left(\scriptstyle 11+28X+22X^2\right)$ \\
			& $\beta _L ^i= \left\{\frac{9}{2}, \frac{3}{4}({ \scriptstyle 1+4X^2 }), {\scriptstyle 5}, \frac{1}{2} \right\}$ &
			$\beta_L ^i= \left\{ \frac{33}{4}, \frac{3}{2} ({\scriptstyle 1+2X+2X^2 }), {\frac{11}{2} }, {\scriptstyle 1},{\scriptstyle 1}, \frac{1}{2}, {\scriptstyle 2} \right\} $\\
			& $\beta_E ^i= \left\{\frac{9}{4},{ \scriptstyle 3(1+X^2)},\frac{5}{2}, {\scriptstyle 1} \right\}$&
			$\beta_E ^i= \left\{ {\frac{9}{4}}, {\scriptstyle 3( 2+2X+X^2)}, {\scriptstyle 3} ,{\frac{5}{2}}, {\frac{3}{2}},{\scriptstyle 1},{\scriptstyle 6} \right\} $  \\
			& &  $\beta_{\kappa}^i = \left\{ \frac{33}{4}, \frac{15}{4}{\scriptstyle +9X+6X^2}, \frac{1}{2} ,\frac{1}{4}, \frac{17}{2}, {\scriptstyle 1}, {\scriptstyle 1} \right\} $\\
			& & $\beta_{\lambda} ^i = \left\{ \scriptstyle 12,10 \right\} $\\
			\bottomrule
		\end{tabular}
	}
	\caption{$\beta$-function coefficients of the RGEs in Eqs.~(\ref{eq:s4RGEIgen}) and (\ref{eq:s4RGEIIgen}), where $\beta _g ^{\text{SM}} = -\frac{19}{3}$ and $\beta _ {g'} ^{\text{SM}} = \frac{41}{3}$. The upper index $i$ of $\beta^i_X$ refers to its components as ordered in Eqs.~(\ref{eq:s4RGEIgen}) and (\ref{eq:s4RGEIIgen}); for example, in model I, $\beta _L ^i =  \{\beta _L ^g, \beta_L ^{g'}, \beta_L ^{L}, \beta_L ^{y}\}$.}
	\label{table:beta_functions}
\end{table}

\clearpage
\section{Kinematics of \boldmath{$2\rightarrow3$} scattering processes}
\label{app:kinematics}

\begin{figure}[!htb]
	\begin{center}
		\includegraphics[width = 0.6\textwidth]{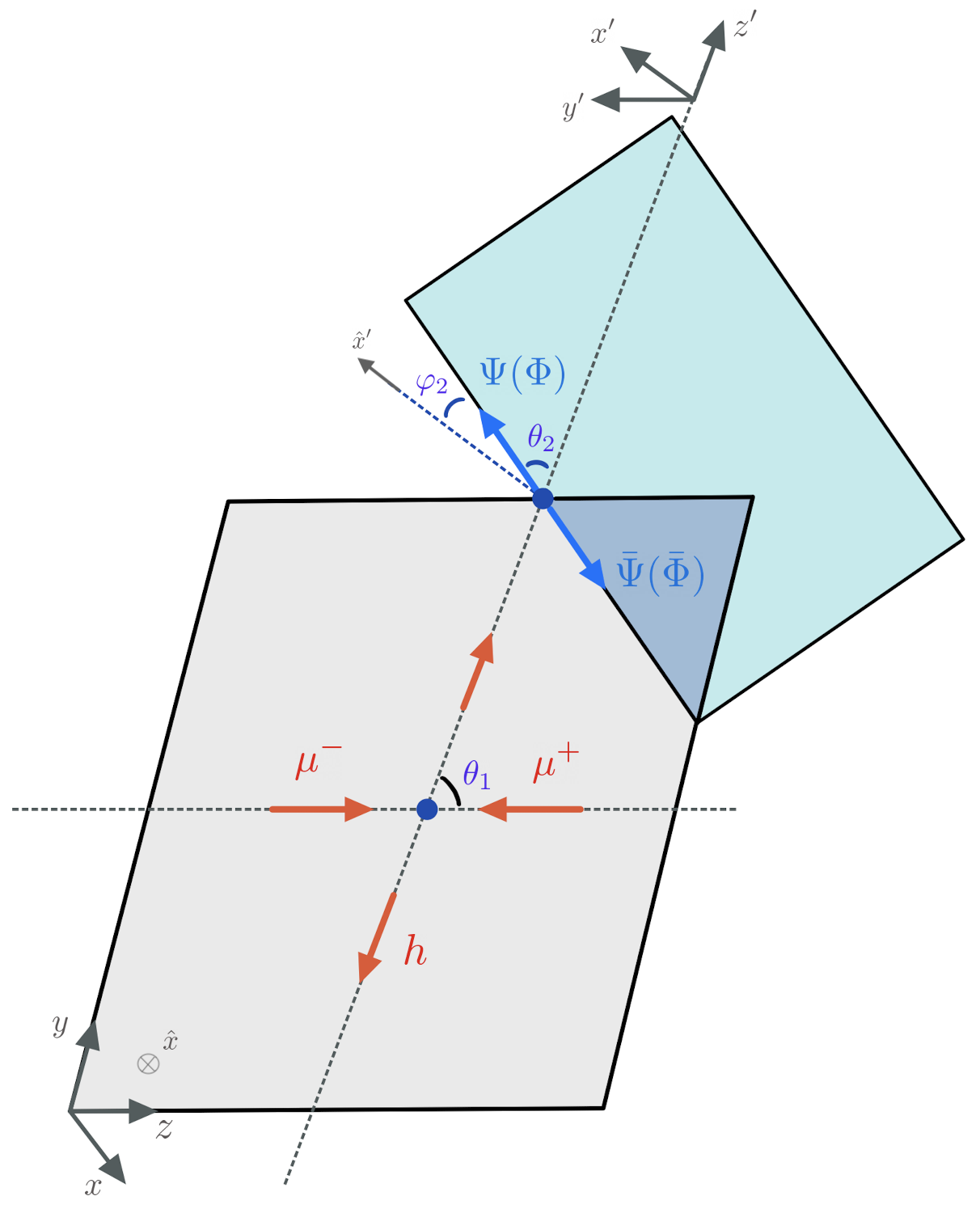}
		\caption{\label{fig:appBframes}The two frames in which the two-body phase-spaces are evaluated. The grey plane is orthogonal to $\hat{x}$ and it is defined in the center-of-mass frame, whereas the blue frame is defined with primed coordinates ${\hat{x}',\hat{y}',\hat{z}'}$ in the rest-frame of the compound particle $q$.}
	\end{center}
\end{figure}

As discussed in section~\ref{section:2_to_3}, the processes $ \mu ^+ \mu ^-\rightarrow h \bar \Psi \Psi$ and $\mu ^+ \mu ^-\rightarrow h \bar \Phi \Phi$
are ideal tests of the muon $g$-2 anomaly. In this appendix, we give details about the kinematics of these processes, which are depicted in Fig.~\ref{fig:Feynman_II}.
The integrated cross section,
\begin{align}
	\sigma_{2\rightarrow 3} &= \int \frac{1}{2s} (2\pi)^4 \delta ^{(4)} (p_a+p_b-k-k_1-k_2)  |\overline{\mathcal{A}}|^2 d\Phi_3\,,
\end{align}
can be evaluated by splitting the phase-space into the product of two-body phase spaces,
\begin{align}
	d\Phi_3&=  \int d q^2 (2\pi)^3 d \Phi_2 (p_a+p_b,k,q) d\Phi_2 (q,k_1,k_2)\,,
\end{align}
corresponding to a $2\rightarrow2$ scattering $p_a+ p_b\rightarrow k+q$ and a decay $q\rightarrow k_1 + k_2$, as depicted in Fig.~\ref{fig:appBframes}.
Exploiting Lorentz invariance of $d\Phi_2$, we compute the $2\rightarrow2$ subprocess in the center-of-mass frame of the $2\rightarrow3$ scattering, with the four-vectors defined as,
\begin{align}
	\begin{cases}
	p_a = (\frac{\sqrt s}{2},0,0\frac{\sqrt s}{2})\\[0.25em]
	p_b = (\frac{\sqrt s}{2},0,0,-\frac{\sqrt s}{2}) \\[0.25em]
	q = (E_q,0,|\mathbf{q}|\sin \theta_1,|\mathbf{q}|\cos \theta_1)\\[0.25em]
	k = (E_k,0,-|\mathbf{q}|\sin \theta_1,-|\mathbf{q}|\cos \theta_1) 
	\end{cases}\,,
\label{eq:appBmomenta1}
\end{align}
whereas the $1\to 2$ process is evaluated in the rest frame of the compound particle $q$, with the following parametrization of the four-vectors,
\begin{align}
	\begin{cases}
		q = (q,0,0,0)\\[0.25em]
		k_1 '= (E_{k_1 '},|\mathbf{k_1 '}|\sin \theta_2 \cos \phi_2,|\mathbf{k_1 '}|\sin \theta_2 \sin \phi_2,|\mathbf{k_1 '}|\cos \theta_2) \\[0.25em]
		k_2 ' = (E_{k_2 '},-|\mathbf{k_1 '}|\sin \theta_2 \cos \phi_2,-|\mathbf{k_1 '}|\sin \theta_2 \sin \phi_2,-|\mathbf{k_1 '}|\cos \theta_2) 
	\end{cases}.
\end{align}
In this expression, the energies and the 3-momenta are evaluated in the respective frames. The angles have been chosen in such a way that the decay frame is reached from the $2\rightarrow2$ frame by a simple Lorentz transformation, $R$, consisting of a rotation  followed by a boost along the $\hat x$ axis,
\begin{align}
	R = \begin{pmatrix}
		 E_q /q & 0 & 0 & -|\mathbf{q}|/q\\
		 0 & 1 & 0 & 0\\
		 0 &0 &1&0 \\
		 -|\mathbf{q}|/q & 0 &0 & E_q /q
	\end{pmatrix} \times
\begin{pmatrix}
	1 & 0 & 0 & 0\\
	0 & 1 &0 &0\\
	0&0& \cos \theta_1 & - \sin \theta_1\\
	0&0&\sin \theta_1 & \cos \theta_1
\end{pmatrix}\,,
\end{align}
where $E_q,|\mathbf{q}|$ are defined in Eq.~(\ref{eq:appBmomenta1}). The scalar products appearing in the amplitudes~ \eqref{eq:s4ampI} and \eqref{eq:s4ampII} are then evaluated in the $2\rightarrow2$ frame by applying the $R$ transformation: $k_1 = R\inv k_1^\prime$ and $k_2 = R\inv k_2^\prime$\,. In our parametrization, the integration intervals are given by $\theta_i \in [0,\pi]$, $\phi_i \in [0,2\pi]$ and $q^2 \in [4M^2, s]$, where the muon and Higgs-boson masses have been neglected.

\newpage

\end{document}